\documentclass{IEEEtran}
\usepackage{cite}
\usepackage{amsmath,amssymb,amsfonts}
\usepackage{algorithmic}
\usepackage{graphicx}
\usepackage{textcomp}
\usepackage{graphicx}
\usepackage{hyperref}
\usepackage{subcaption} % Add this in your preamble if not already included
\def\BibTeX{{\rm B\kern-.05em{\sc i\kern-.025em b}\kern-.08em
    T\kern-.1667em\lower.7ex\hbox{E}\kern-.125emX}}

\usepackage{verbatim} % usually available by default
\usepackage{algorithm}
\usepackage{algorithmic}
\usepackage{xcolor}

\DeclareUnicodeCharacter{2218}{\ensuremath{^\circ}}
\DeclareUnicodeCharacter{2218}{\ensuremath{^\circ}}
\DeclareUnicodeCharacter{2212}{-}

\begin{document}
\title{Millimeter-Wave Multi-Radar Tracking System Enabled by a Modified GRIN Luneburg Lens for Real-Time Healthcare Monitoring}
\author{Mohammad Omid Bagheri, \IEEEmembership{Member, IEEE}, Justin Chow, Josh Visser, Veronica Leong, \IEEEmembership{Member, IEEE}, and George Shaker, \IEEEmembership{Senior Member, IEEE}
\thanks{This work was supported in part by the AP-S/URSI 2025 Student Design Contest Organizing Committee and Wireless Sensor and Devices Lab (WSDL) at University of Waterloo, Ontario, Canada}
\thanks{M. O. Bagheri (Postdoc Fellow, omid.bagheri@uwaterloo.ca), V. Leong (M.S. student, vsleong@uwaterloo.ca), J. Chow (Undergraduate student, j63chow@uwaterloo.ca), J. Visser (Undergraduate student, j3visser@uwaterloo.ca), and G. Shaker (Supervisor, gshaker@uwaterloo.ca) are with the Department of Electrical and Computer Engineering, University of Waterloo, Waterloo, ON, Canada.}}

\maketitle

\begin{abstract}
Multi-beam radar sensing systems are emerging as powerful tools for non-contact motion tracking and vital-sign monitoring in healthcare environments. This paper presents the design and experimental validation of a synchronized millimeter-wave multi-radar tracking system enhanced by a modified spherical gradient-index (GRIN) Luneburg lens. Five commercial FMCW radar modules operating in the 58–63 GHz band are arranged in a semi-circular configuration around the lens, whose tailored refractive-index profile accommodates bistatic radar modules with co-located transmit (TX) and receive (RX) antennas. The resulting architecture generates multiple fixed high-gain beams with improved angular resolution and minimal mutual interference. Each radar operates independently but is temporally synchronized through a centralized Python-based acquisition framework to enable parallel data collection and low-latency motion tracking. A 10-cm-diameter 3D-printed prototype demonstrates a measured gain enhancement of approximately 12 dB for each module, corresponding to a substantial improvement in detection range. Full-wave simulations and measurements confirm effective non-contact, privacy-preserving short-range human-motion detection across five 28$^\circ$ sectors, providing 140$^\circ$ total angular coverage. Fall-detection experiments further validate reliable wide-angle performance and continuous spatial tracking. The proposed system offers a compact, low-cost, and scalable platform for millimeter-wave sensing in ambient healthcare and smart-environment applications.
\end{abstract}


\section{Introduction}
\IEEEPARstart{T}{he} increasing demand for advanced healthcare monitoring systems has driven substantial progress in intelligent sensing and monitoring technologies aimed at improving patient safety, early diagnosis, and continuous health assessment. Recent developments have enabled real-time, non-contact, and privacy-preserving operation, making such systems particularly suitable for clinical and home-based environments. These technologies have been successfully applied across a broad range of medical domains, including cardiac health monitoring \cite{gharamohammadi2025smart}, remote monitoring of oncology patients \cite{closs2024application}, breast cancer detection \cite{bhatti2025advances,hernandez2025icabme}, respiratory-disorder assessment \cite{tao2023clinical}, and vital signs detection \cite{chen2024high}. Among these, fall detection, particularly for elderly individuals and patients with mobility impairments, has emerged as a critical research focus due to its direct impact on healthcare outcomes and quality of life \cite{hu2024radar}.

%%%%%%%%%%%%%%%%%%%%%%%%%%%%%%%%
\begin{figure}[!t]
\centerline{\includegraphics[trim=420 600 300 450, clip, width=\linewidth]{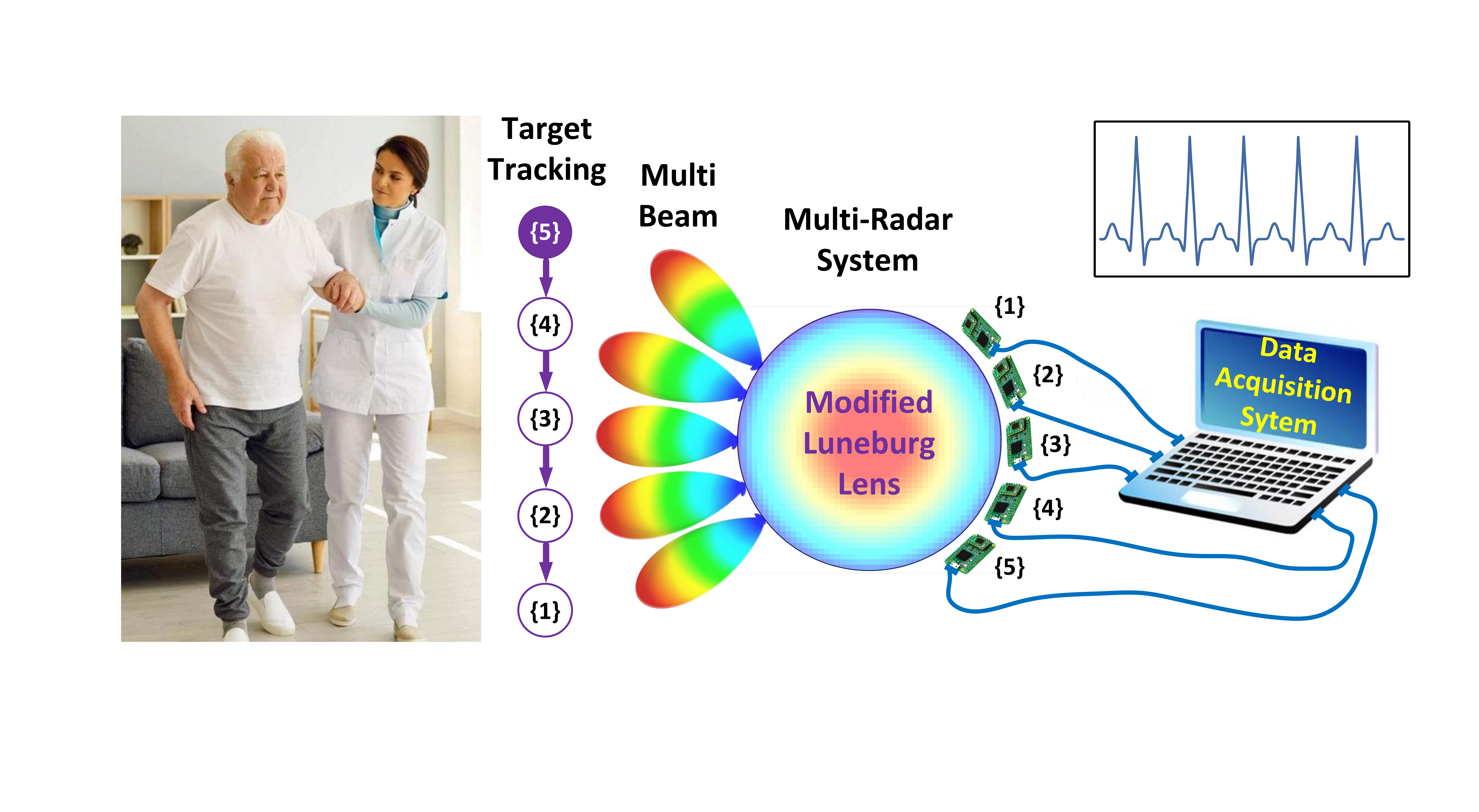}}
\caption{Configuration of the synchronized multi-radar tracking system for real-time, non-contact healthcare monitoring.}
\label{fig1}
\vspace{-0.3cm}
\end{figure}
%%%%%%%%%%%%%%%%%%%%%%%%%%%%%%%%%%

To meet the growing demand for accurate and unobtrusive health monitoring, bistatic radar-on-chip technologies have emerged as scalable, compact, and cost-effective solutions. In particular, millimeter-wave (mm-wave) radar systems have gained considerable attention due to their ability to provide high spatial resolution, real-time responsiveness, and suitability for integration into wearable or ambient sensing platforms. Such systems have been increasingly employed in applications requiring precise motion detection and physiological monitoring, including autonomous navigation, industrial inspection, gesture recognition, and various biomedical scenarios \cite{soumya2023recent}. Within the healthcare domain, mm-wave radar technology has enabled advanced sensing capabilities for fall detection \cite{hall2024validating}, gait analysis \cite{abedi2022hallway}, and non-invasive glucose monitoring \cite{bagheri2024radar,bagheri2024metasurface}, demonstrating its versatility for continuous and privacy-preserving human monitoring.

However, single-radar systems often provide insufficient spatial coverage and transmit power for reliable patient monitoring, particularly in real-world healthcare environments. To enhance coverage and tracking accuracy, advanced beamforming capabilities are essential. Traditional mechanical beam steering methods rely on physically rotating antennas or reflectors, offering directional scanning but suffering from slow response time, high complexity, and mechanical wear \cite{montaser2021deep}. Electronic beam steering using phased arrays and MIMO architectures provides faster response and flexible control but often requires costly phase shifters, complex feed networks, and precise calibration—factors that limit scalability and accessibility in compact healthcare systems \cite{cardillo2020review}. To address these limitations, \cite{yuan2021recent} introduced a metamaterial leaky-wave antenna for non-contact vital-sign detection, achieving frequency-controlled beam steering at 5.8 GHz and detection up to 2 m. Similarly, a 60 GHz multi-layer leaky-wave antenna in \cite{mingle2023multi} demonstrated a 24 dB gain, ~30$^\circ$ scanning range, and respiration/heartbeat detection up to 4 m. While compact and real-time, both designs rely on frequency-dependent beam steering, limiting simultaneous multi-target tracking and reducing efficiency in wideband or fast monitoring scenarios.

Multi-beam architecture, as an alternative approach, has been introduced in wireless communications \cite{hong2017multibeam} to enable simultaneous beam formation in multiple directions without mechanical or electronic steering. These passive structures provide instantaneous, frequency-independent angular coverage and generate multiple fixed beams, resulting in lower cost, reduced hardware complexity, and enhanced robustness. This approach offers improved spatial diversity, wide angular coverage, and high radiation efficiency, making it particularly well-suited for continuous, real-time monitoring of human motion and physiological activity.

In radar sensing systems, multi-beam operation can be achieved using either a single-radar multi-beam configuration or a multi-radar distributed architecture. In the single-radar approach, a single feed is integrated with a lens or metasurface to passively split the radiated power into multiple beams, enabling simultaneous observation of different directions \cite{bagheri2017dual,bagheri2025dual}. In contrast, the multi-radar configuration employs multiple spatially distributed radar modules that operate in a synchronized manner to collectively expand spatial coverage and increase effective gain. While single-radar multi-beam systems offer compactness and simplicity, they inherently suffer from reduced efficiency and limited beam control due to passive power division. Conversely, multi-radar fusion architectures provide superior spatial coverage, higher SNR, improved robustness, and greater flexibility, making them a more effective solution for real-time, high-resolution healthcare monitoring and motion tracking applications \cite{shen2023indoor}. 

The use of multi-radar fusion techniques has facilitated a range of sensing applications, including multibin respiration analysis in driver monitoring systems using 60-GHz FMCW radars \cite{gharamohammadi2023multibin}, distributed radar-based human gait classification \cite{li2020sequential}, continuous motion detection \cite{zhang2025distributed}, indoor multi-target tracking \cite{he2025drmtrack}, and collision prediction for road safety \cite{choi2021road}. In \cite{schellenberger2020continuous}, a four-radar 77 GHz FMCW setup mounted beneath a mattress enabled continuous patient monitoring with over 95\% activity recognition accuracy and a 40\% reduction in false detections compared to single-radar systems. Similarly, \cite{iwata2021multiradar} demonstrated a dual 79 GHz radar system for non-contact respiratory monitoring of multiple individuals via data fusion. In \cite{bagheri2025multi}, a multi-radar configuration operating in the 58–63 GHz band was used for near-field physiological sensing, utilizing spatial diversity to improve parameter detection on human skin.

An emerging strategy in multi-radar domains is the intentional use of overlapping radar beams or fields of view, ensuring that adjacent sensing regions reinforce one another and preventing targets from being lost as they move across sector boundaries. \cite{luo2024distributed} show that distributed phased-MIMO radars form their surveillance region through overlapping TX/RX beam volumes, enabling controlled spatial resolution. Similarly, the SuperDARN network relies on overlapping fields of view across 35+ stations to maintain spatial continuity \cite{mcwilliams2023borealis}, and national S-band/C-band weather radar networks use jointly illuminated regions for robust coverage \cite{lee2021real}. Although these examples illustrate the value of multi-radar overlap, they are fundamentally designed for long-range, low- or mid-frequency sensing and involve large-scale infrastructures or multistatic MIMO architectures. 

At the same time, the broader multi-radar fusion literature shows that coherent or even consistent data integration requires strict temporal alignment and unified processing pipelines \cite{jinming2022time}. In the mm-wave domain, reviews of radar-based perception and multi-object tracking report similar challenges: asynchronous FMCW radars often generate range–Doppler maps with unaligned sampling, inconsistent chirp timing, and differing noise statistics, making sector-to-sector comparison unreliable without coordinated acquisition \cite{soumya2023recent,qian2025review}. Additionally, distributed radar studies indicate that MIMO geometry and antenna placement strongly influence SNR, angular resolution, and coverage uniformity \cite{wang2022geometric}. Together, these findings highlight the necessity of temporal synchronization, consistent signal processing, and robust fusion strategies in any multi-radar sensing system.

Despite extensive work on multi-radar networks at HF, VHF, and S-band frequencies, and a growing body of research on mm-wave sensing for perception and environmental monitoring, a clear gap remains in the development of compact, short-range, high-resolution mm-wave systems capable of continuous angular sensing using multiple commercial FMCW radars with real-time data fusion. This gap becomes even more pronounced in biomedical contexts, where full-angle patient motion tracking introduces additional design and coordination challenges. Specifically, such systems must be arranged along a semi-circular path to ensure overlapping fields of view, which in turn introduces issues such as mutual interference and signal coupling between closely spaced radar modules, effects known to degrade detection reliability \cite{wang2014radar}. Furthermore, achieving improved radar-on-chip performance often requires the development of custom antenna structures, which in turn necessitates a complete redesign and fabrication of the radar module. This approach significantly increases cost and complexity, making it impractical for applications that demand flexible or application-specific performance. Consequently, an external beam-forming element is required to enhance angular coverage and control radiation characteristics without modifying the underlying radar hardware.

Lens technologies have been widely employed to enhance radiation efficiency and amplify both transmitted and received signals, thereby improving the SNR and overall detection sensitivity. In designing a multi-radar system, where each radar module operates in a bistatic configuration with distinct TX/RX antennas arranged on the same plane, a key challenge lies in selecting an appropriate lens architecture capable of managing the complex radiation behavior across multiple radiating elements \cite{bagheri2025dielectric}. An integrated grooved Fresnel lens operating at 24 GHz has been demonstrated for system-on-package implementations incorporating a fractal antenna array \cite{ghaffar201124}. Similarly, a dual-layer PCB-based planar lens array developed for 77 GHz automotive radar applications achieved a 15 dB gain enhancement, while a comparable design for 60 GHz near-field biomedical sensing improved radar SNR by 13 dB \cite{bagheri2024radar}. Despite these advantages, these lens configurations are not well-suited for multi-radar configurations, as they limit independent beam control and increase the risk of inter-radar interference.

Gradient-index (GRIN) lenses offer an attractive solution for multi-radar sensing systems due to their high gain, beam-steering capability, low-profile design, and cost-effective fabrication. Their symmetric spherical geometry with a gradient dielectric profile ensures polarization independence, making them particularly suitable for configurations where radar modules are distributed along a semi-circular arc \cite{bagheri2025uniformly}. With multiple feeds positioned around the lens, each feed generates a distinct, well-collimated beam in a specific direction, enabling true multi-beam operation \cite{li2019multibeam}. When combined with lens-based focusing, as illustrated in Fig. \ref{fig1}, the system achieves enhanced angular resolution and improved robustness to occlusion, key features for real-time applications such as fall detection in clinical and assisted-living environments.

To implement practical multi-radar architectures, bistatic mm-wave radar modules such as the Infineon BGT60TR13C are widely adopted due to their compact form factor, integrated TX and RX channels, and high modulation bandwidth. These features make them particularly suitable for applications in remote outdoor environments and real-time healthcare monitoring \cite{markovic202560,bagheri2024radar,bagheri2024metasurface}. However, a challenge in integrating such radar modules with a conventional Luneburg lens (LL) lies in feed-induced beam tilting, which can distort the intended radiation direction and degrade system performance. When the transmit or receive antennas are displaced from the ideal focal geometry, the resulting radiation beam deviates from the boresight direction, leading to reduced gain, beam misalignment, and degraded SNR within the intended field of view. In \cite{bagheri2025dielectric}, an X-band radar module integrated with a Luneburg lens exhibited a beam tilt of approximately 13$^\circ$, primarily caused by feed offset from the optimal focal position. To mitigate this effect, a correction technique employing dual-beam masks was introduced to locally modify the phase distribution and restore on-axis beam alignment, thereby recovering the nominal boresight gain and SNR performance. The modified Luneburg Lens (MLL)-enhanced radar module presented in \cite{bagheri2025dielectric} achieved a 10 dB enhancement in both realized gain and received power at 10.4 GHz, corresponding to a 3.2-fold increase in detection range and a tenfold improvement in SNR when operating along the boresight.

To the best of available literature, the integration of a modified GRIN Luneburg lens within a synchronized multi-radar architecture has not been previously reported. This paper proposes passive beamforming structure serves to shape, focus, and spatially distribute the transmitted and received energy, as well as addressing issues such as mutual interference, synchronization drift, and non-uniform spatial coverage that commonly arise in closely spaced multi-radar configurations. The complete system implements five commercial 58–63 GHz FMCW radar modules arranged along a semi-circular path around the multi-radar MLL to produce continuous angular sectors that collectively achieve 140$^\circ$ spatial coverage. The radars are temporally synchronized through a centralized Python-based acquisition framework that ensures aligned chirp timing and uniform sampling across all sensors, enabling reliable cross-radar comparison. A dedicated graphical user interface (GUI) is developed in parallel to support real-time visualization of patient motion and to provide immediate fall-detection alerts. The resulting system enables continuous spatial scanning, allowing a patient to move naturally between adjacent sensing sectors while maintaining reliable detection and tracking. 

The main contributions of this work are summarized as follows: (i) Introduction of a Multi-Radar Modified Luneburg Lens (MMLL) Architecture: Unlike prior classical or modified Luneburg lenses, the GRIN architecture proposed in this work is specifically designed for multi-radar operation. A rod-based, direction-dependent modification to the permittivity profile creates multiple independent high-gain beam channels aligned to five bistatic mm-wave radar modules, representing the first GRIN lens capable of simultaneous multi-beam operation without mechanical or electronic steering. (ii) Full-System Demonstration for Real-Time Fall Detection: this work presents a complete mm-wave radar sensing platform that integrates multi-beam generation, synchronized signal acquisition, and real-time processing for dynamic patient monitoring, including fall detection. The system is designed for clinical and assisted-living settings, enabling wide-angle and continuous observation of human activity. Unlike conventional depth cameras, wearable devices, or single-beam radars, the proposed architecture provides unobtrusive, contact-free monitoring that preserves patient privacy. (iii) Multi-Beam Functionality with Wide-Angle Coverage: the lens-based architecture provides over $140\circ$ of angular sensing coverage through spatial beam distribution, eliminating the need for active beam steering. This makes the system well suited for low-power, low-cost, real-time monitoring applications where continuous angular awareness is required. (iv) High Sensing Functionality: the proposed system provides more than 12 dB of experimentally measured realized-gain enhancement for the lens-integrated transmit and receive antennas, which corresponds to a theoretical fourfold extension in detection range relative to standalone FMCW radars. This enhanced link budget substantially improves detection resolution and reliability across all sensing directions. (v) Modular, Compact, and Scalable Hardware Platform: the system is constructed using low-cost 3D-printing technology to realize the GRIN lens structure and incorporates compact off-the-shelf radar modules. This modular design enables rapid prototyping, simplified fabrication, and scalable deployment across diverse sensing environments.

%%%%%%%%%%%%%%%%%%%%%%%%%%%%%%%%%%%%%%%%%%%

\section{Multi-Radar Modified GRIN Luneburg Lens (MMLL)}

Arranging multiple radar modules along a semicircular path for multi-beam sensing inherently introduces strong mutual coupling and beam distortion, making a standalone configuration impractical. The GRIN Luneburg lens is of considerable interest for integration with antenna systems due to its inherent beam-steering capabilities and high-gain performance. When paired with a feeding antenna, the lens enables directional radiation that can be effectively analyzed using ray-tracing methods. A conventional GRIN Luneburg lens exhibits a radially inhomogeneous permittivity profile derived from a refractive index distribution, assuming a constant magnetic permeability. This spatial variation in dielectric permittivity allows electromagnetic waves to bend and focus as they propagate through the lens. When a source antenna is placed at the lens surface, the radially decreasing permittivity guides the rays toward the opposite side, resulting in highly directive radiation. The classic permittivity profile for such a lens can be expressed as,

\begin{equation}
n(r) = \sqrt {{\varepsilon _r}.{\mu _r}}  = \sqrt {2 - {{(\frac{r}{R})}^2}}
\label{eq1}
\end{equation}

\hspace{-0.335cm}where $r = \sqrt{x^2 + y^2 + z^2}$ denotes the radial distance from the center of the sphere, ranging from $0$ to $\textit{R}$, with $x$, $y$, and $z$ representing the spatial coordinates, and $\textit{R}$ indicating the external radius of the spherical lens~\cite{guo2020optimal}.

\begin{figure}[!t]
\centering
\begin{subfigure}{\columnwidth}
    \centering
    \includegraphics[width=0.95\linewidth]{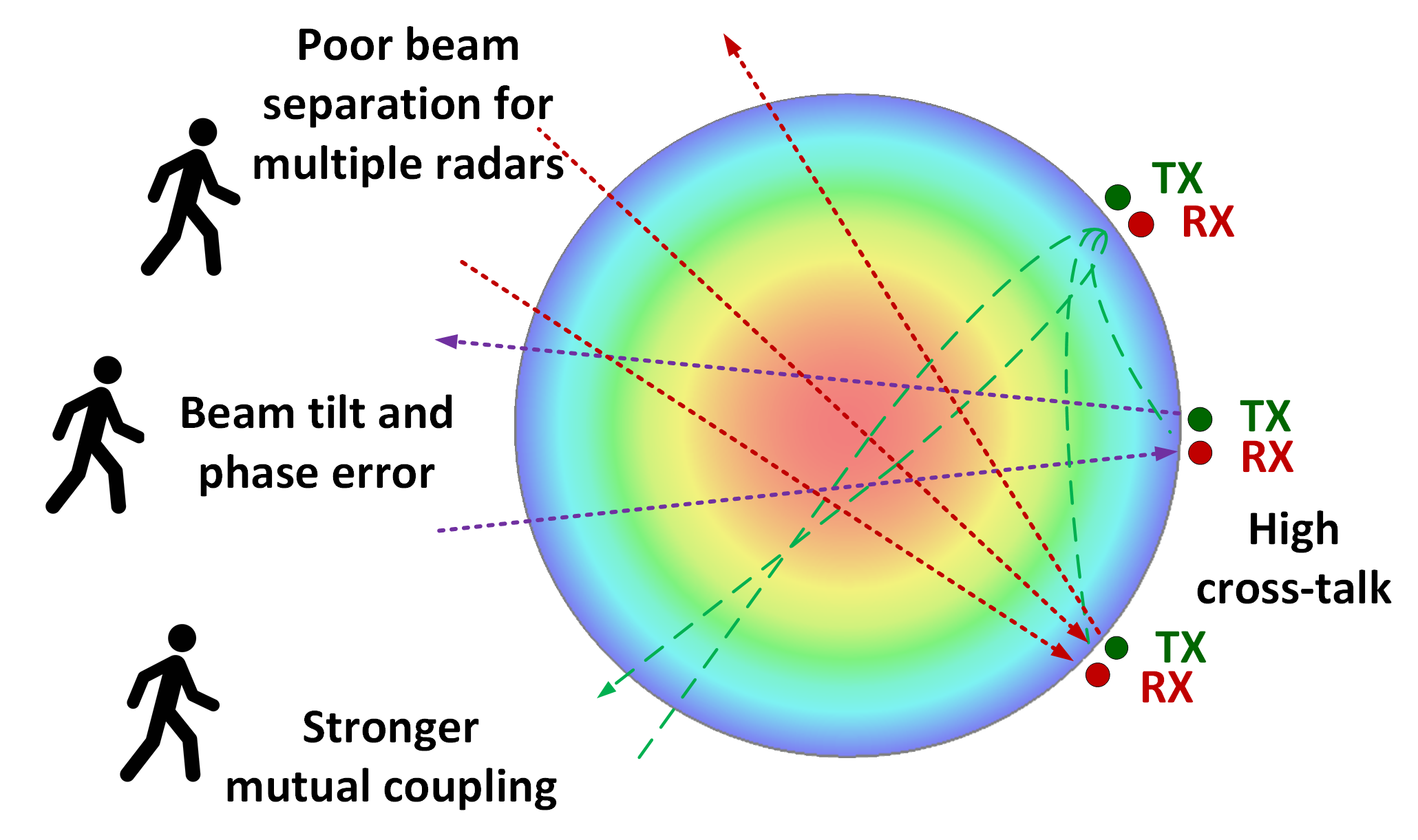}
    \caption{}
\end{subfigure}
\begin{subfigure}{\columnwidth}
    \centering
    \includegraphics[width=0.95\linewidth]{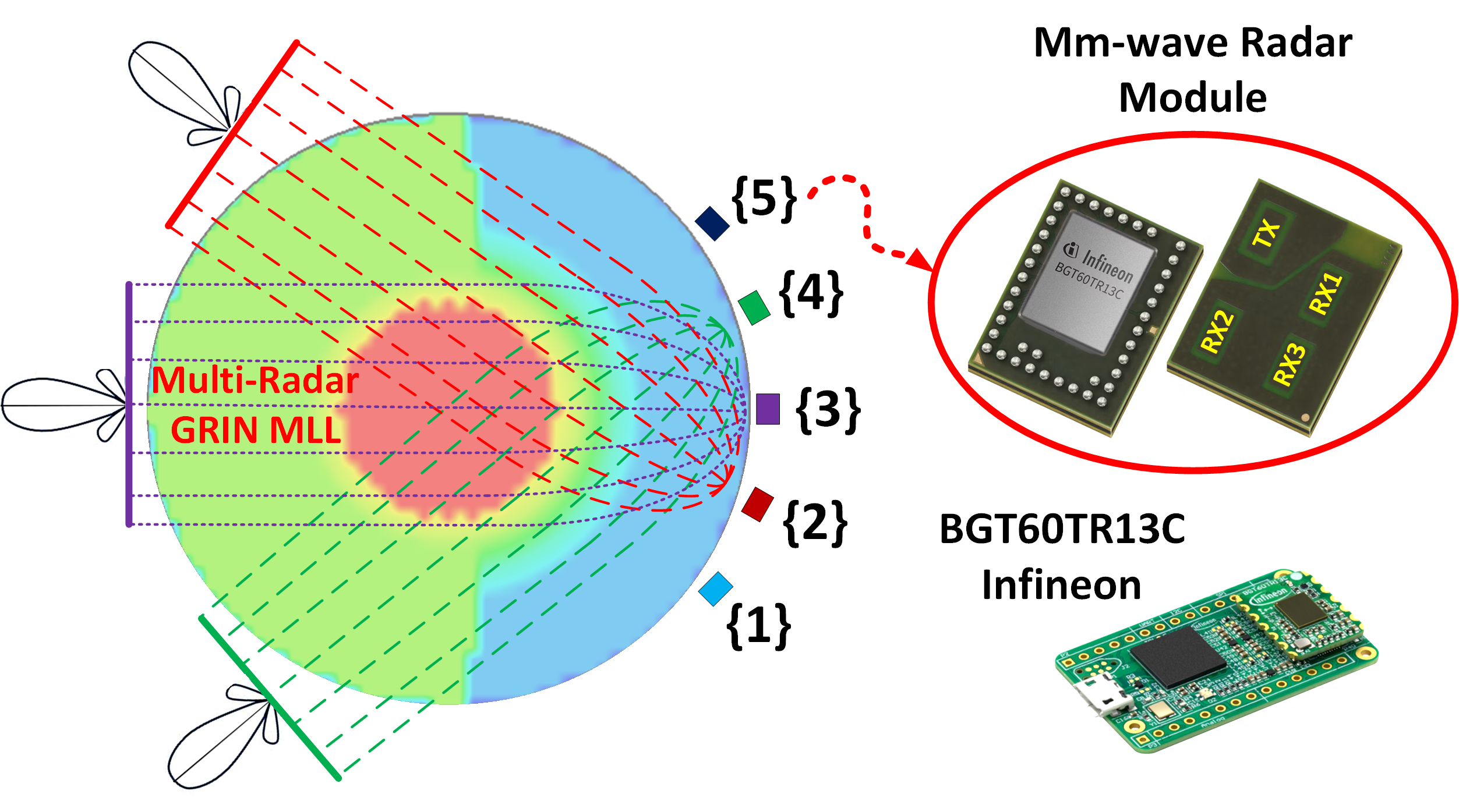}
    \caption{}
\end{subfigure}
\caption{(a) illustration of the performance degradation caused by the classical Luneburg permittivity profile when used with multiple bistatic radar modules. (b) Ray-tracing configuration of the required permittivity distribution in the modified Luneburg lens, with five Infineon BGT60TR13C radar modules.}
\label{fig2}
\vspace{-1.5mm}
\end{figure}

Conventional Luneburg lens designs typically employ a single, centrally positioned feed to achieve symmetric excitation and a focused boresight beam. To expand the field of view for tracking applications, however, multiple feeds must be integrated into the lens structure. When an on-chip bistatic radar module is used as a feed, as shown in Fig.~\ref{fig2}(a), each antenna operates as an independent off-axis radiator. The resulting lateral displacement between the TX and RX elements introduces phase deviations from the ideal GRIN distribution, distorting wavefront formation and producing beam tilt, aperture asymmetry, and misalignment between the transmit and receive main lobes. These effects reduce realized gain, often below 15 dB, degrade SNR within the sensing region \cite{bagheri2025dielectric}, and create unequal TX/RX performance across radar modules. Furthermore, the unmodified Luneburg index profile cannot compensate for the off-focal feed positions, nor can it suppress the increased mutual coupling and cross-talk that arise when multiple radars operate simultaneously. As a result, the classical lens exhibits overlapping or de-focused beams, elevated sidelobe levels, and insufficient angular separation, collectively diminishing detection accuracy and compromising overall system robustness. These limitations underscore the need for a modified permittivity profile capable of supporting multi-radar integration while preserving high efficiency for each unit independently, as demonstrated in Fig.~\ref{fig2}(b).

%%%%%%%%%%%%%%%%%%%%%%%%%%%%%%%%
%%%%%%%%%%%%%%%%%%%%%%%%%%%%%%%%%%

In this section, the development and validation of the Multi-Radar Modified GRIN Luneburg Lens are detailed, serving as the passive beamforming core for wide-angle, multi-directional mm-wave radar sensing. The MMLL is designed to support multiple bistatic radar modules arranged along its periphery, enabling simultaneous high-gain beam generation with minimal cross-interference. The theoretical design methodology and engineered permittivity distribution used to construct the modified lens structure, along with the corresponding simulation and measurement results that validate beam behavior, gain performance, and sensing coverage of the fabricated prototype, are presented in the following.

%%%%%%%%%%%%%%%%%%%%%%%%%%%%%%%%%%%%%

\subsection{MMLL Design Theory}

The development of the MMLL is initially inspired by our previous work in \cite{bagheri2025dielectric}, which introduced a modified GRIN Luneburg lens (MLL) incorporating strategically placed dielectric perturbation elements within the lens structure to locally adjust the phase distribution. This approach merges a tailored refractive index gradient with computationally optimized dielectric masks, enabling control of beam direction and focal location. The design effectively mitigates the phase errors caused by offset-fed antennas, realigning both the TX and RX beams toward the boresight direction ($\theta$ = 0$^\circ$). Consequently, it significantly enhances beam alignment, gain consistency, and SNR. 

%%%%%%%%%%%%%%%%%%%%%%%%%%%%%%%%
\begin{figure}[!t]
\centering
\begin{subfigure}{0.5\columnwidth}
    \centering
    \includegraphics[width=\linewidth]{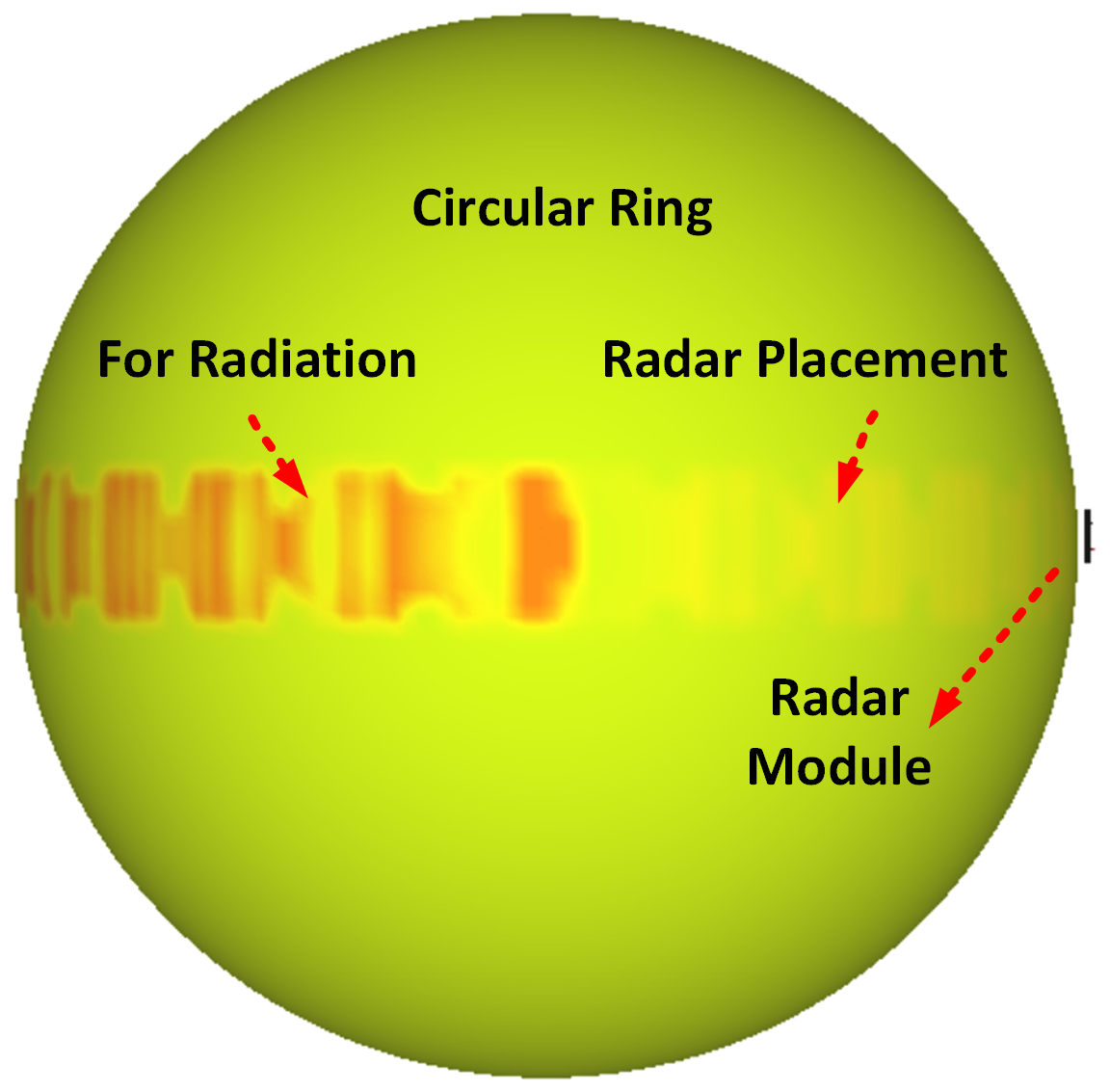}
    \caption{}
\end{subfigure}
\hfill
\begin{subfigure}{0.4\columnwidth}
    \centering
    \includegraphics[width=\linewidth]{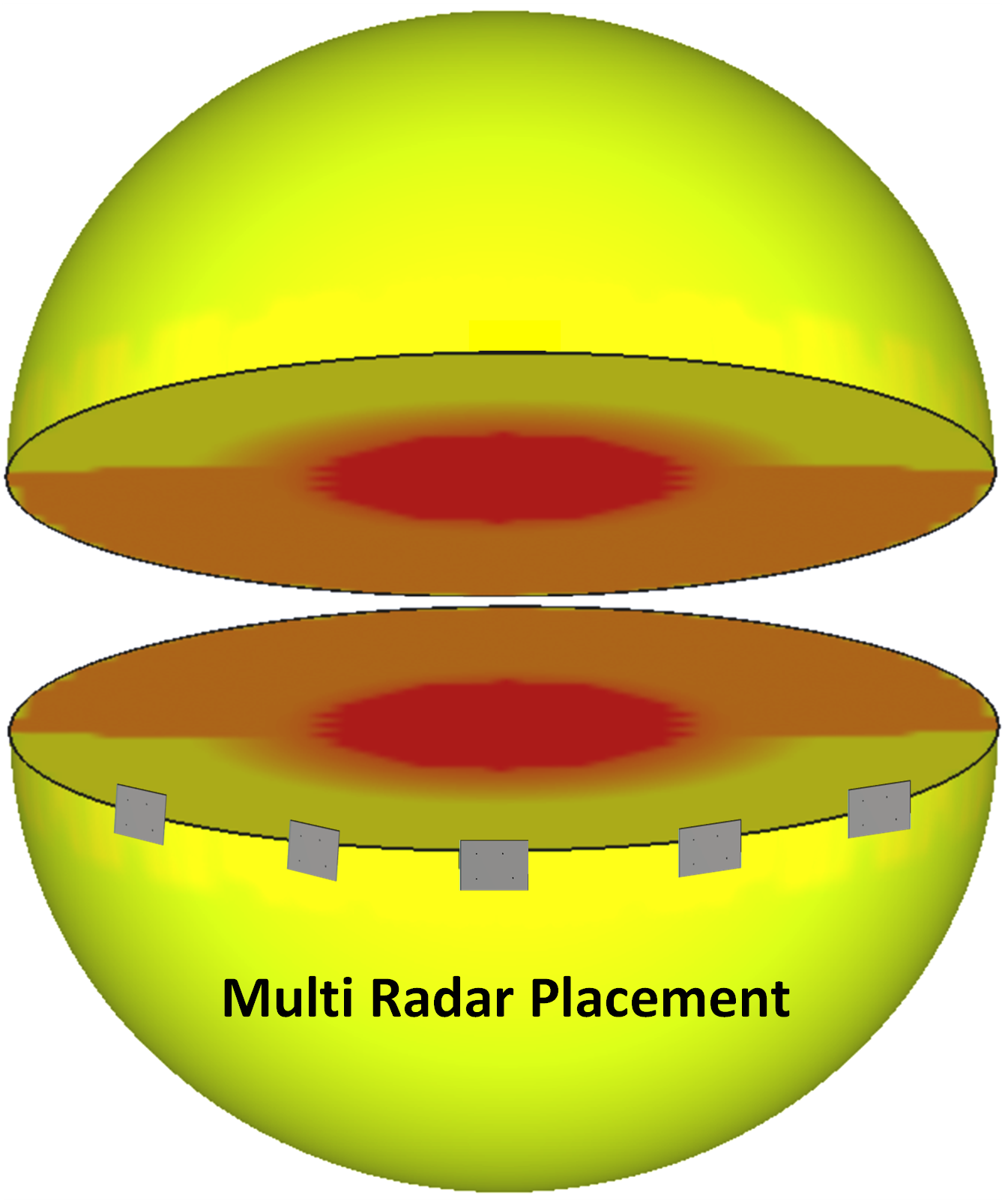}
    \caption{}
\end{subfigure}
\caption{(a) Full spherical view of the modified Luneburg lens with a circular ring-based layout for integrating radar modules, highlighting designated regions for beam radiation and radar placement. (b) Cross-sectional view of the lens showing the permittivity distribution tailored for a multi-radar configuration to enable distributed sensing.}
\label{fig3}
\end{figure}
%%%%%%%%%%%%%%%%%%%%%%%%%%%%%%%%%%

%%%%%%%%%%%%%%%%%%%%%%%%%%%%%%%%
\begin{figure}[!t]
\centering
\begin{subfigure}{0.48\columnwidth}
    \centering
    \includegraphics[width=\linewidth]{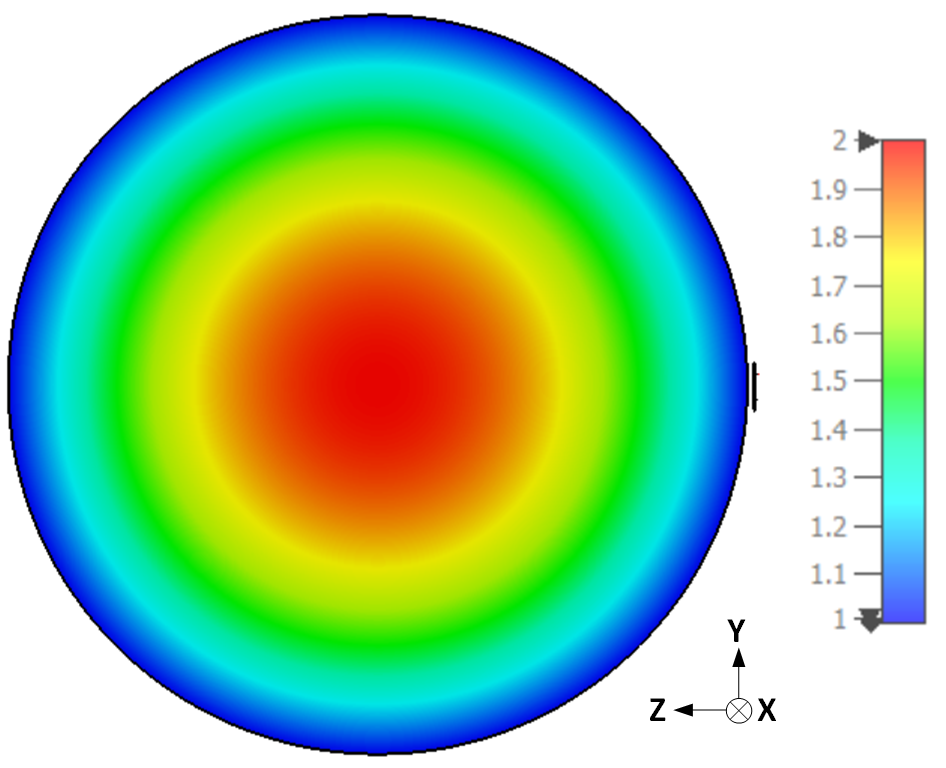}
    \caption{}
\end{subfigure}
\hfill
\begin{subfigure}{0.48\columnwidth}
    \centering
    \includegraphics[width=\linewidth]{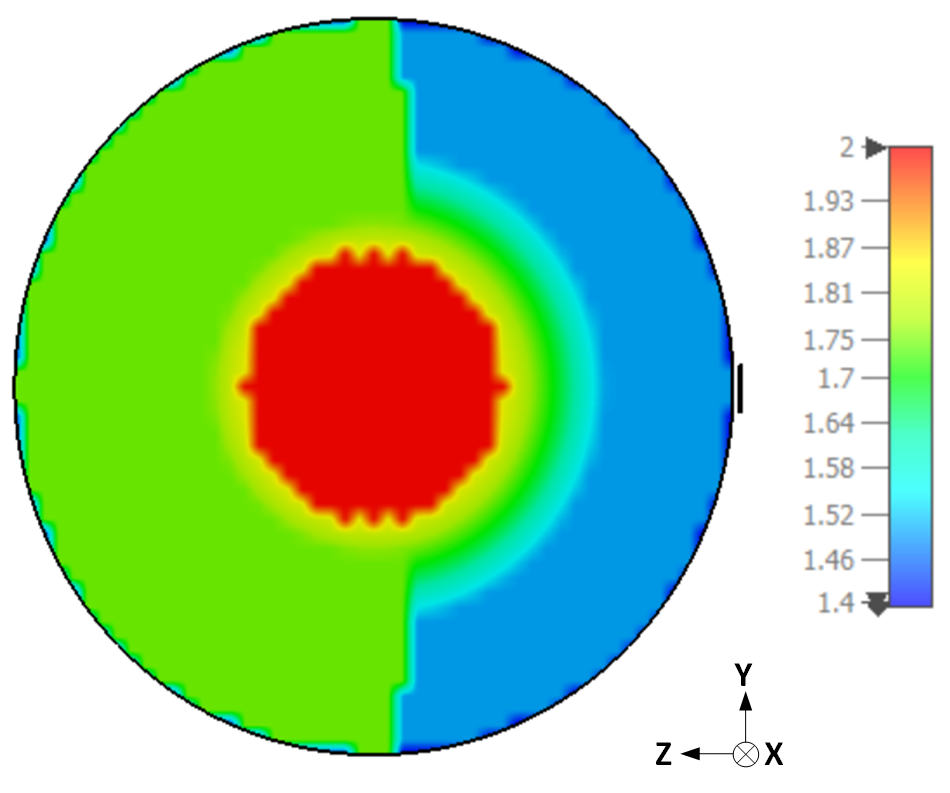}
    \caption{}
\end{subfigure}
\hfill
\begin{subfigure}{0.48\columnwidth}
    \centering
    \includegraphics[width=\linewidth]{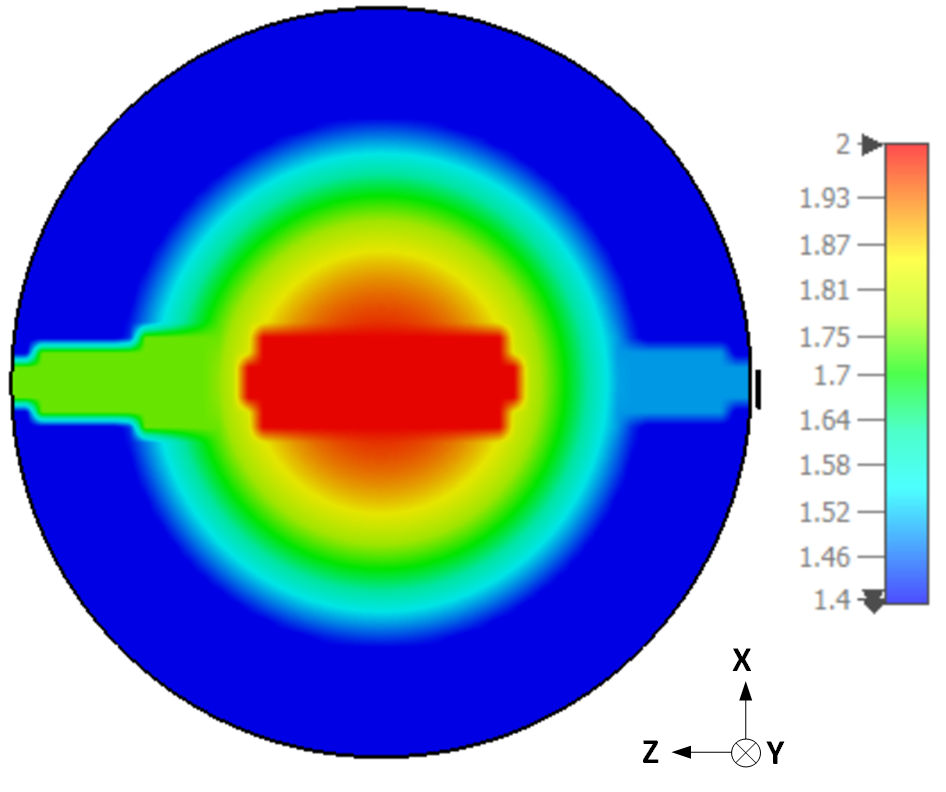}
    \caption{}
\end{subfigure}
\hfill
\begin{subfigure}{0.48\columnwidth}
    \centering
    \includegraphics[width=\linewidth]{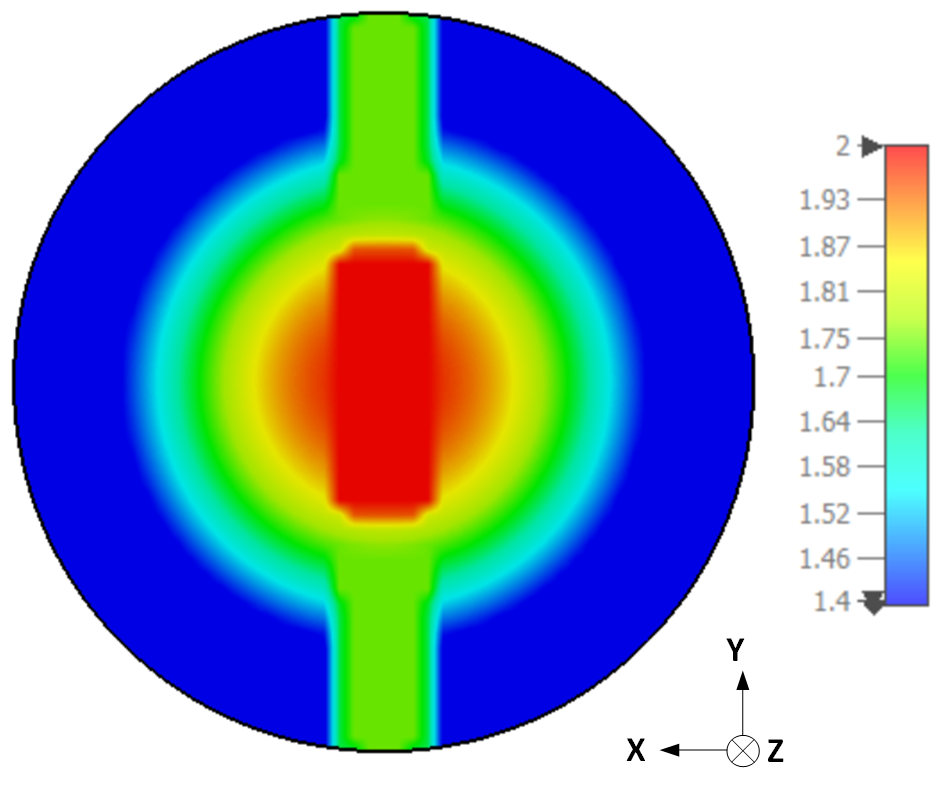}
    \caption{}
\end{subfigure}
\caption{(a) Conventional permittivity distribution of a Luneburg lens with radius R. (b) 2D cross-section of the modified Luneburg lens in the X–Z plane, illustrating the 180$^\circ$-rotated dual-beam mask configuration—where the dark-blue region represents radar module placement and the green hemisphere indicates the radiation side. (c) Cross-section in the Y–Z plane. (d) Cross-section in the X–Y plane.}
\label{fig4}
\end{figure}
%%%%%%%%%%%%%%%%%%%%%%%%%%%%%%%%%%

Despite these improvements, applying the previously developed MLL design to mm-wave applications presents several challenges. First, the much shorter wavelengths at mm-wave frequencies demand extremely high fabrication precision, as even sub-millimeter inaccuracies in the dielectric mask or gradient profile can introduce significant phase errors and degrade beamforming performance. In addition, the dual-beam mask optimization technique, effective for radar alignment at lower frequencies, does not scale efficiently for radar configurations at mm-wave, where angular alignment is more sensitive and beam control must be more precise. Furthermore, the original design was tailored for single-radar use and does not address the complexities introduced by integrating multiple TX/RX radar modules, including mutual interference and the need for simultaneous multi-beam alignment.

Compared with classical and previously modified Luneburg lenses, the proposed GRIN architecture introduces several fundamental advances that enable true multi-radar operation. Classical LLs employ a symmetric r/R permittivity profile, as shown in Fig. \ref{fig2}(a), with a single focal point and are therefore intrinsically suited only for single-feed excitation; they cannot compensate for the asymmetric TX/RX illumination of bistatic radar modules or prevent beam distortion when multiple radars are used. Prior modified LLs, including dual-beam mask and localized phase-correction designs \cite{bagheri2025dielectric}, address beam tilt for individual radars but remain single-feed solutions that do not mitigate mutual coupling or support simultaneous multi-beam formation. In contrast, the present design incorporates a rod-based, direction-dependent GRIN modification in which multiple dielectric rods, each comprising three longitudinal permittivity segments, form a continuous anisotropic region across the feed, facing hemisphere. Although the rods overlap rather than correspond to specific feeds, each radar illuminates a different effective cross-section of this anisotropic region, producing the localized phase redistribution required to suppress feed-induced beam tilt and stabilize beam formation. The rod arc also provides angularly selective dielectric loading that confines energy within the intended sector of each radar, thereby reducing mutual coupling and minimizing spillover into neighboring fields of view. Together with the asymmetric dual-region structure separating the radar-placement and radiation zones, this architecture effectively transforms the classical single focal point into a distributed focal arc tailored to the geometry of the multiple bistatic radar modules, enabling wide-angle, high-gain, multi-beam operation using a single compact spherical aperture—capabilities not supported by prior LL or GRIN implementations.

Building upon the previous methodology, the new modified lens features a GRIN permittivity distribution with the rotationally symmetric rod-based scheme, implemented as multiple overlapping masks, is introduced to form distinct beam paths arranged in a semi-circular configuration, with each path tailored to a specific radar node, as shown in Fig. \ref{fig3}a and \ref{fig3}b. This architecture is optimized for mm-wave sensing applications, specifically targeting the Infineon BGT60TR13C radar modules operating in the 58–63 GHz band. To achieve this, the permittivity distribution of the spherical lens with radius \( R \) is re-engineered by inserting a rotational dielectric rod into the conventional GRIN index profile, Fig.~\ref{fig4}(a), where the rod is swept about the x-axis to occupy one side of the sphere, thereby distinguishing the radar-feed side from the radiation-emission side. This modification produces the revised GRIN distribution illustrated in Fig.~\ref{fig4}(b)–(d) for the 2D cross-sections in the X–Z, Y–Z, and X–Y planes, respectively. This configuration realizes a rod-based synthesis scheme that supports multiple beam paths distributed in a semi-circular configuration. As depicted in Fig. \ref{fig4}, the rods fully overlap to form a continuous anisotropic region rather than distinct isolated channels, ensuring that each feed interacts with a different effective cross-section of this structure depending on its angular placement.

The presented permittivity distribution offers distinct advantages for multi-radar systems owing to its asymmetric GRIN profile, which strategically separates the radar placement and radiation regions. The rod structure comprises three distinct sections with constant permittivity values ranging from 1.4 to 2.0, arranged symmetrically and rotated by $180^\circ$, forming two concentric semi-ring regions and a central circular core. As illustrated in Fig. \ref{fig4}(b), the blue region, characterized by the lowest permittivity ($\varepsilon_r = 1.4$), is designated for radar placement to achieve impedance matching between the radar aperture and free space. Although an ideal match would require $\varepsilon_r = 1$, fabrication limitations at mm-wave frequencies restrict the minimum achievable value, introducing a small phase deviation that slightly reduces overall efficiency. The green-to-red portion of the lens exhibits a higher permittivity gradient that effectively guides and focuses radiated energy outward, ensuring unidirectional transmission and improved beam symmetry across different feeds. The central red region, with the highest permittivity ($\varepsilon_r = 2.0$), enhances field confinement and directs the radiated phase fronts toward the desired propagation direction. 

The last region, shown in green, corresponds to the radiation side of the lens. Using a constant-permittivity layer in this region, rather than extending the classical Luneburg gradient down to the lower-permittivity blue zone (($\varepsilon_r \approx 1.4$)) for ideal impedance matching, represents a deliberate design trade-off. This choice prioritizes beam shaping and field-focusing performance over perfect boundary matching. The uniform higher-permittivity layer helps prevent beam skewing caused by asymmetric off-axis feeds, acts as a stabilizing buffer for the outgoing wavefront, reduces phase sensitivity to feed-position variations, suppresses mutual coupling among adjacent radar modules, and supports consistent high-gain beam formation across all sensing units, while also simplifying fabrication and integration.

The proposed modified GRIN Luneburg lens supports simultaneous radar access from multiple angles, offering a scalable and efficient solution for multi-angle distributed sensing platforms. To reproduce the modified permittivity distribution shown in Fig.~\ref{fig4} and map it onto the Luneburg lens geometry, a computational procedure is defined to evaluate the permittivity at each spatial point \((x, y, z)\) as follows:

\vspace{-1em}

\begin{equation}
\varepsilon_r(x, y, z) = \max \left\{ \varepsilon_{\min},\ \varepsilon_{\text{GRIN}}(x, y, z),\ \max_{i=1}^{N} \psi_i(x, y, z) \right\}
\label{eq:total_eps}
\end{equation}

\noindent where \(\varepsilon_{\min} = 1.38\) ensures a minimum physical permittivity, and \(\varepsilon_{\text{GRIN}}(x, y, z)\) is the base GRIN profile:
    \begin{equation}
    \varepsilon_{\text{GRIN}}(x, y, z) = 0.8 + \left(1 - \frac{x^2 + y^2 + z^2}{R^2} \right) \cdot 1.2
    \end{equation}
\(\psi_i(x, y, z)\) accounts for the contribution of the $i$-th rod. Each rod is aligned along a unit vector \(\vec{v}_i\) placed symmetrically in a half-ring configuration within the Y-Z plane:

\begin{equation}
\vec{v}_i =
\begin{bmatrix}
0 \\
\sin \theta_i \\
\cos \theta_i
\end{bmatrix}, \quad \theta_i \in [-90^\circ, +90^\circ],\quad i = 1, \dots, N
\end{equation}

\noindent For a point \(\vec{p} = [x, y, z]^T\), we define:
\begin{align}
\ell_i(x, y, z) &= \vec{p} \cdot \vec{v}_i \quad \text{(projection length)} \\
\vec{c}_i(x, y, z) &= \ell_i \cdot \vec{v}_i \quad \text{(closest point on rod axis)} \\
r_{\perp,i}^2(x, y, z) &= \| \vec{p} - \vec{c}_i \|^2 \quad \text{(squared radial distance)}
\end{align}

\noindent The permittivity contribution of each rod is defined piecewise:

\begin{equation}
\psi_i(x, y, z) = 
\begin{cases}
\varepsilon_{r1}, & \text{if } \ell_i \in [-R, -0.34R],\ r_{\perp,i}^2 < r_{\text{tube}}^2 \\
\varepsilon_{r2}, & \text{if } \ell_i \in (-0.34R, +0.34R],\ r_{\perp,i}^2 < r_{\text{tube}}^2 \\
\varepsilon_{r3}, & \text{if } \ell_i \in (+0.34R, +R],\ r_{\perp,i}^2 < r_{\text{tube}}^2 \\
0, & \text{otherwise}
\end{cases}
\end{equation}

\noindent where: \(R = 50\, \text{mm}\) is the lens radius, \(r_{\text{tube}} = 7.5\, \text{mm}\) is the fixed radius of each rod, \(\varepsilon_{r1} = 1.5,\ \varepsilon_{r2} = 2,\ \varepsilon_{r3} = 1.75\) are the permittivity values assigned in the lower, middle, and upper sections of each rod, respectively.

This formulation enables the core structure follows a classical GRIN profile, with permittivity values varying between 1.4 and 2, ensuring beam-focusing capability, while embedded rods along curved paths, (180$^\circ$ arc) around the X-axis, introduce directional energy channels to accommodate multiple input/output radar beams. The orientation of each rod is defined by a vector \(\vec{v}_i\) spanning from $-90^\circ$ to $+90^\circ$ in the Y-Z plane, Fig. \ref{fig4}, and is segmented into three axial zones with distinct permittivity values to enable controlled phase shifting. The "max" operator is applied across the GRIN and rod-defined permittivities to retain the dominant dielectric value at each spatial coordinate.

This design approach is scalable across a wide range of frequencies and radar module configurations; however, the present work is specifically tailored to meet the operational requirements of mm-wave sensing systems, particularly those utilizing commercial Infineon BGT60TR13C radar modules operating within the 58–63 GHz band, as shown in Fig. \ref{fig1}. In this configuration, each radar module generates a fixed, high-gain beam, and collectively, these beams form multiple synchronized phase fronts that provide continuous spatial coverage without the need for mechanical or electronic beam steering. The modified GRIN Luneburg lens architecture thereby enables simultaneous radar access from multiple angular positions, laying the foundation for advanced multi-angle, distributed mm-wave sensing applications.

The rod-based dielectric modification provides several key advantages for multi-beam and multi-feed operation. Instead of assigning one rod to each radar, the overlapping semi-ring arrangement of rods collectively introduces a direction-dependent anisotropy into the GRIN profile. When different radars illuminate the lens from different angular positions, each feed interacts with a different effective cross-section of this anisotropic region, resulting in localized phase adjustments that stabilize its beam and reduce feed-induced tilt. This shared anisotropic structure also redistributes energy flow inside the lens in a way that naturally limits spillover between adjacent angular sectors, thereby reducing mutual coupling and cross-interference among the five radar modules. As a result, the lens behaves not as a single symmetric GRIN profile but as a multi-feed-optimized aperture with angularly varying phase correction, enabling simultaneous formation of multiple independent high-gain beams, a capability not achievable using classical or symmetrically modified Luneburg lenses.

%%%%%%%%%%%%%%%%%%%%%%%%%%%%%%%%%%%%%%%%%%%%%%%%%%%%%%%%%%

\subsection{Validation of modified Luneburg Lens Performance}

The proposed tracking system is a multi-radar configuration employing five Infineon 60~GHz radar modules integrated with a modified spherical Luneburg lens. These radar modules are selected for their compact form factor, mm-wave operation, and capability to perform high-resolution imaging and precise motion detection. Strategically positioned in a 2D planar arrangement, each module is oriented to cover a distinct angular sector, enabling wide-area monitoring. At the core of the system, the designed Luneburg lens facilitates efficient beam focusing and steering, enhancing directional control and signal gain across multiple radar feeds.

All radar modules are connected to a central processing unit responsible for data acquisition, synchronization, and signal processing. This unit employs advanced algorithms to extract key signal parameters such as magnitude, phase, and range. The processed data enables localization and motion tracking of targets within the monitored area. A Python-based software interface manages system control and provides real-time data visualization and analytics. The interface displays detailed power plots received from each radar module and incorporates algorithms to enhance the SNR and minimize false detections.

\begin{figure}[!t]
\centerline{\includegraphics[width=\columnwidth]{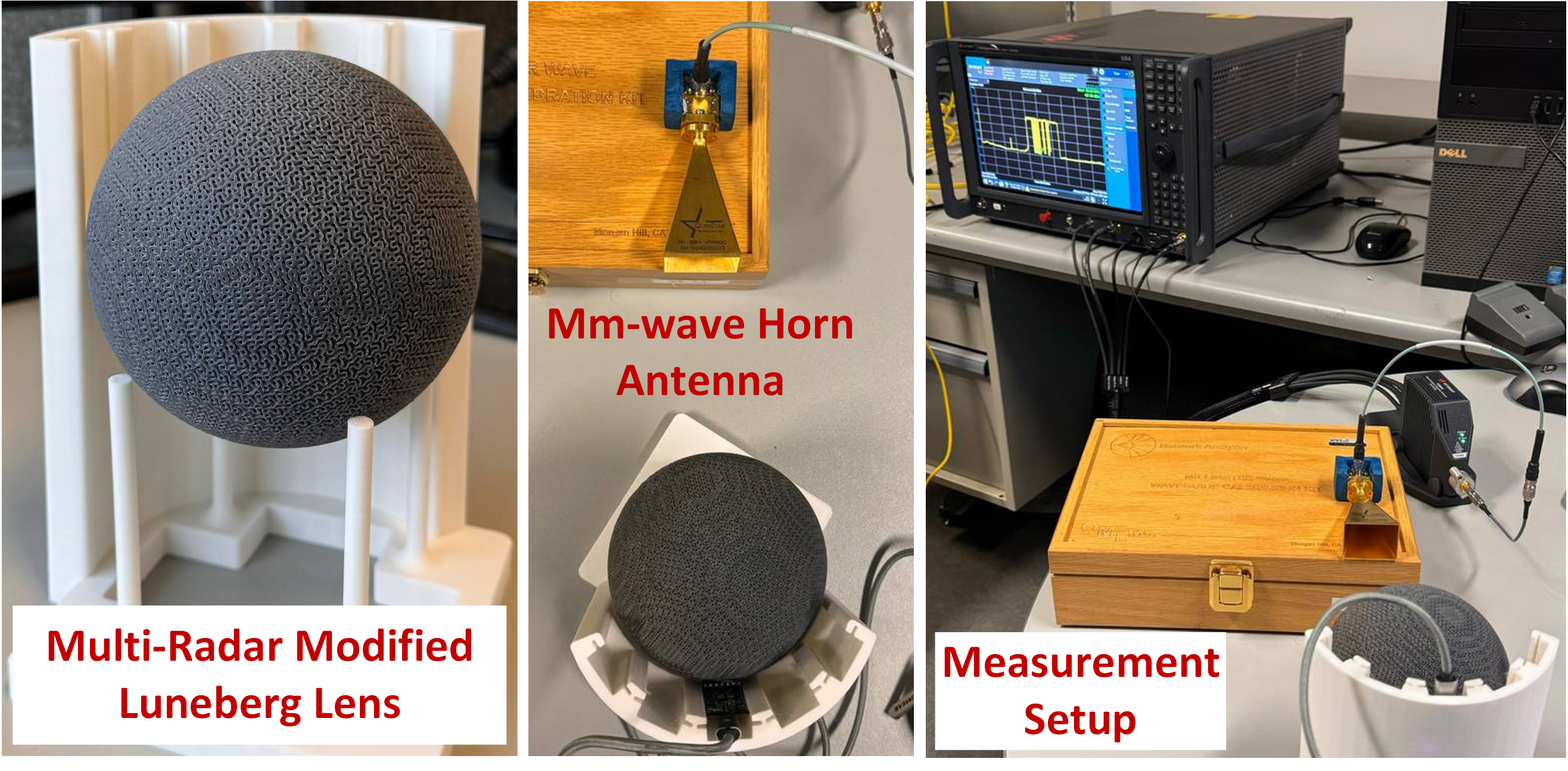}}
\caption{Fabricated prototype of the modified Luneburg lens integrated with a radar module, with a mm-wave horn antenna connected to a spectrum analyzer for transmitted power measurement.}
\label{fig5}
\end{figure}

Before proceeding to the tracking system evaluation, it is necessary to assess the performance of the proposed Luneburg lens, both through full-wave simulations and experimental measurements, to validate its ability to enhance transmission power when integrated with five Infineon 60-GHz radar modules arranged around the lens with an angular spacing of 28$^\circ$. Fig.~\ref{fig5} shows a fabricated prototype of the proposed spherical multi-radar modified Luneburg lens, featuring a 10-cm diameter. The presented prototype utilizes additive manufacturing technology, providing flexibility in both design and fabrication, thereby streamlining the realization of GRIN materials. In this work, Fortify’s digital light processing (DLP) additive manufacturing system is utilized to achieve precise and efficient lens fabrication \cite{hobart2023novel}. Specifically, the Fortify Flux Core DLP 3D printer is employed to produce the spherical GRIN-based Modified Luneburg Lens (MLL) using Rogers Corporation’s Radix™ resin, a 3D-printable dielectric with a relative permittivity of 2.8 and a low loss tangent of 0.0043. This material enables single-exposure layer printing, ensuring high-resolution structural integrity and reduced fabrication time.

%%%%%%%%%%%%%%%%%%%%%%%%%%%%%%%%
%%%%%%%%%%%%%%%%%%%%%%%%%%%%%%%%%%

The printable design is derived from the simulation data generated by Algorithm \ref{alg1}, Adapted from (2) to (8), which is then translated into a 3D gyroid-based structure representing the spatial permittivity distribution. In artificial dielectric media, the size of the unitcell dictates both the maximum operational frequency and manufacturing cost. The proposed gyroid unitcell, forming the fundamental element of the lens, has a cubic geometry with a 2 mm side length, the smallest value achievable under Fortify’s current printing capabilities. This dimension minimizes phase error deviations in the GRIN profile and supports an optimal performance up to 60 GHz. When the operating wavelength approaches the unitcell size, the effective-medium approximation deteriorates, leading to reduced antenna gain. Considering the target frequency range of 58–63 GHz, corresponding to a center wavelength of approximately 5 mm, the 2 mm unitcell remains close to the theoretical limit, (around $0.4\lambda$), which is expected to decrease permittivity continuity and thus lower lens efficiency.

A custom 3D-printed fixture is employed to securely position the modules, aligning each radar antennas to radiate toward the center of the spherical lens along the circular ring layout, as depicted in Fig.~\ref{fig3}a. This arrangement enables each module to cover a distinct angular sector of approximately 28$^\circ$ within a 2D plane, ensuring comprehensive spatial coverage. All radar modules are interfaced with a central processing unit tasked with managing data acquisition, synchronization, and signal processing.

%%%%%%%%%%%%%%%%%%%%%%%%%%%%%%%%%%%%%%%%%%
\begin{algorithm}[t]
\caption{Multi-Beam Modified Luneburg Lens Generation}
\label{alg1}
\begin{algorithmic}

\STATE \textbf{Input:}
\STATE \hspace{\algorithmicindent} $R \leftarrow$ Radius of the Luneburg lens
\STATE \hspace{\algorithmicindent} $r_{\text{step}} \leftarrow$ Step size for spatial discretization
\STATE \hspace{\algorithmicindent} $r_t \leftarrow$ Base radius of dielectric rods
\STATE \hspace{\algorithmicindent} $N_{\text{rods}} \leftarrow$ Number of symmetric rods
\STATE \hspace{\algorithmicindent} $\theta_i \in [-90^{\circ}, 90^{\circ}] \leftarrow$ Angular directions for each rod

\STATE \textbf{Output:}
\STATE \hspace{\algorithmicindent} $\varepsilon(x,y,z) \leftarrow$ 3D modified permittivity matrix
\STATE \hspace{\algorithmicindent} ASCII file $\leftarrow$ Fabrication data for voxel-based import

\vspace{4pt}
\STATE \textbf{Grid Generation:}
\STATE Discretize $x, y, z$ over spherical domain using $r_{\text{step}}$
\STATE Create 3D mesh: $X, Y, Z \leftarrow \text{meshgrid}(x, y, z)$

\vspace{4pt}
\STATE \textbf{Compute Base GRIN Profile:}
\STATE $\varepsilon_{\text{base}} \leftarrow 0.8 + \left(1 - \dfrac{X^2 + Y^2 + Z^2}{R^2}\right) \times 1.2$

\vspace{4pt}
\STATE \textbf{Generate Symmetric Rod Directions:}
\FOR{$i = 1$ to $N_{\text{rods}}$}
    \STATE Compute unit vector $\mathbf{v}_i = [0, \sin(\theta_i), \cos(\theta_i)]$
\ENDFOR

\vspace{4pt}
\STATE \textbf{Modify Permittivity for Multi-Beam Formation:}
\FOR{each rod direction $\mathbf{v}_i$}
    \STATE Project grid points: $L = Xv_x + Yv_y + Zv_z$
    \STATE Compute perpendicular distance: $r_{\perp}^2 = (X,Y,Z)^2 - L^2$
    \STATE DEFINE three longitudinal segments:
    \STATE \hspace{1.8em} Segment 1: $L \in [-50, -17] \Rightarrow \varepsilon = 1.5$
    \STATE \hspace{1.8em} Segment 2: $L \in [-17, 17] \Rightarrow \varepsilon = 2$
    \STATE \hspace{1.8em} Segment 3: $L \in [17, 50] \Rightarrow \varepsilon = 1.75$
    \STATE APPLY $\varepsilon_{\text{dual}} = \max(\varepsilon_{\text{dual}}, \varepsilon_{\text{updated}})$
\ENDFOR
\STATE CLIP $\varepsilon_{\text{dual}} < 1.38 \Rightarrow 1.38$

\vspace{4pt}
\STATE \textbf{Save Fabrication Data:}
\STATE FLATTEN $(x, y, z, \varepsilon_{\text{dual}})$ into a 2D array
\STATE EXPORT as ASCII file for CST import and 3D printing

\vspace{4pt}
\STATE \textbf{Visualization:}
\STATE GENERATE cross-sectional plots for:
\STATE \hspace{1.8em} $Z=0$ (X–Y plane), $Y=0$ (X–Z plane), and $X=0$ (Y–Z plane)
\STATE DISPLAY permittivity maps using color scale for validation

\end{algorithmic}
\end{algorithm}
%%%%%%%%%%%%%%%%%%%%%%%%%%%%%%%%%%%%%%%%%%%%%%

\begin{figure}[!t]
\centering

\subfloat[]{\includegraphics[width=\columnwidth]{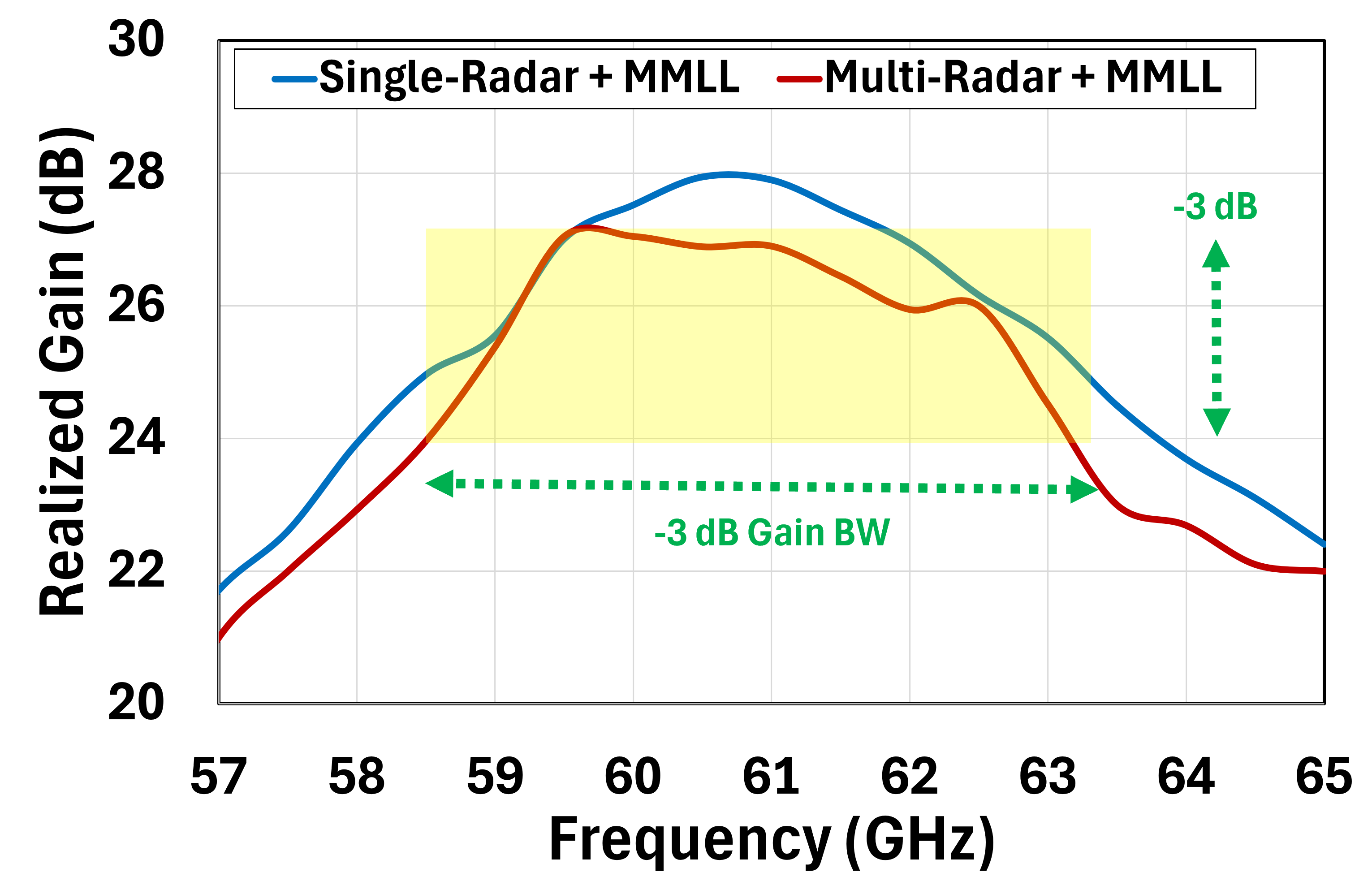}}\label{fig6a}\hfill
\subfloat[]{\includegraphics[width=\columnwidth]{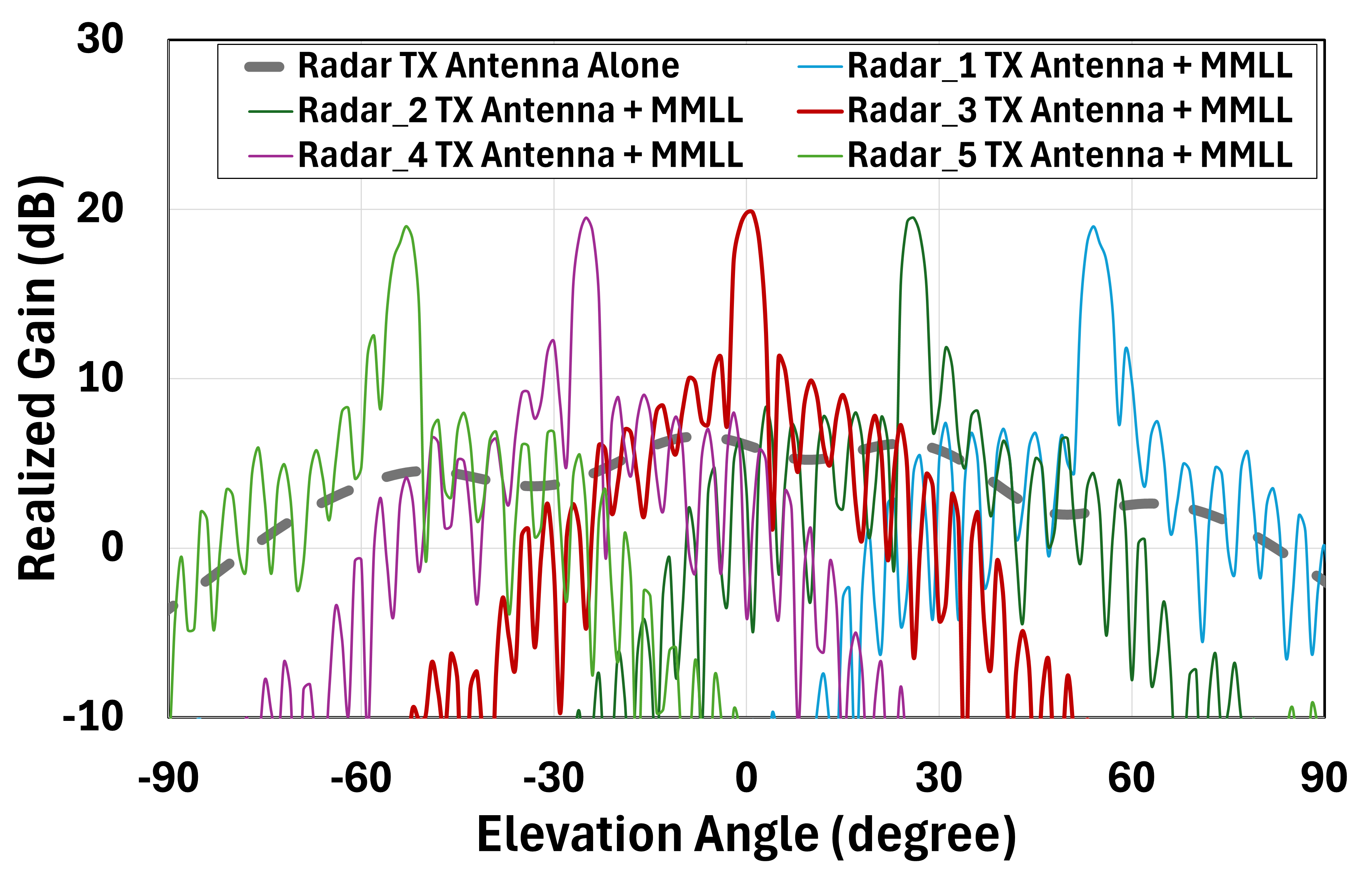}}\label{fig6b}
\subfloat[]{\includegraphics[width=\columnwidth]{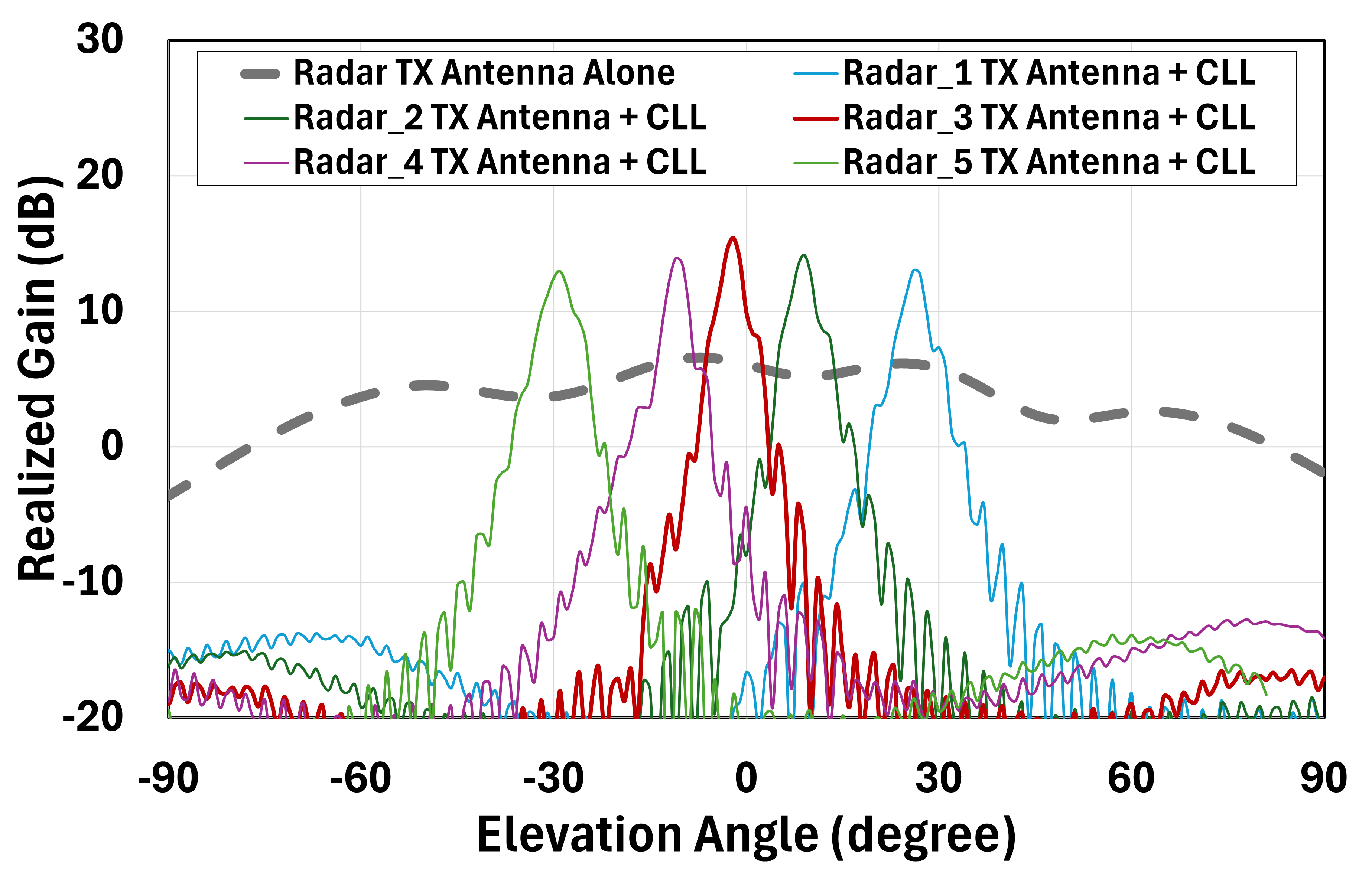}}\label{fig6c}

\caption{Transmission power analysis: (a) Simulated realized-gain variation over frequency for single-radar and multi-radar operation in the presence of the modified multi-feed Luneburg lens (MMLL). (b) Simulated realized gain comparison with and without the modified lens (MMLL). (c) Simulated realized gain comparison with and without the conventional Luneburg lens (CLL).}
\label{fig6}
\end{figure}

To validate the lens performance, both its ability to enhance transmitted power for higher radiation gain and its ability to improve received power, directly influencing the SNR, must be evaluated. As shown in Fig.~\ref{fig6}(a), the realized-gain response of a single radar operating with the MMLL is compared to the case where all five radars operate simultaneously. In the single-radar condition, the peak realized gain reaches approximately 28 dB near 60 GHz. Under multi-radar operation, the peak gain remains high at approximately 27 dB, with a corresponding -3 dB gain bandwidth spanning 59–63 GHz. The small reduction of 1 dB in peak gain, attributable to shared aperture effects and minor mutual interactions, and the preservation of the 4-GHz gain bandwidth confirm that the MMLL maintains broadband radiation characteristics even when all five radars operate concurrently. The close agreement between the two gain–frequency curves verifies that the proposed MMLL effectively mitigates inter-radar interference and ensures stable, high-gain beam performance for each module within the 58–63 GHz band. 

However, the fabrication constraints on the minimum achievable unitcell size can introduce non-uniformities in the permittivity distribution, resulting in higher phase errors that must be carefully accounted for during design optimization.

%%%%%%%%%%%%%%%%%%%%%%%%%%%%%%%%%%%%%%%%%%%%%%%%%

%%%%%%%%%%%%%%%%%%%%%%%%%%%%%%%%%%%%%%%%%%%%%%%

%%%%%%%%%%%%%%%%%%%%%
\begin{figure}[t!]
  \centering
  \includegraphics[width=\linewidth]{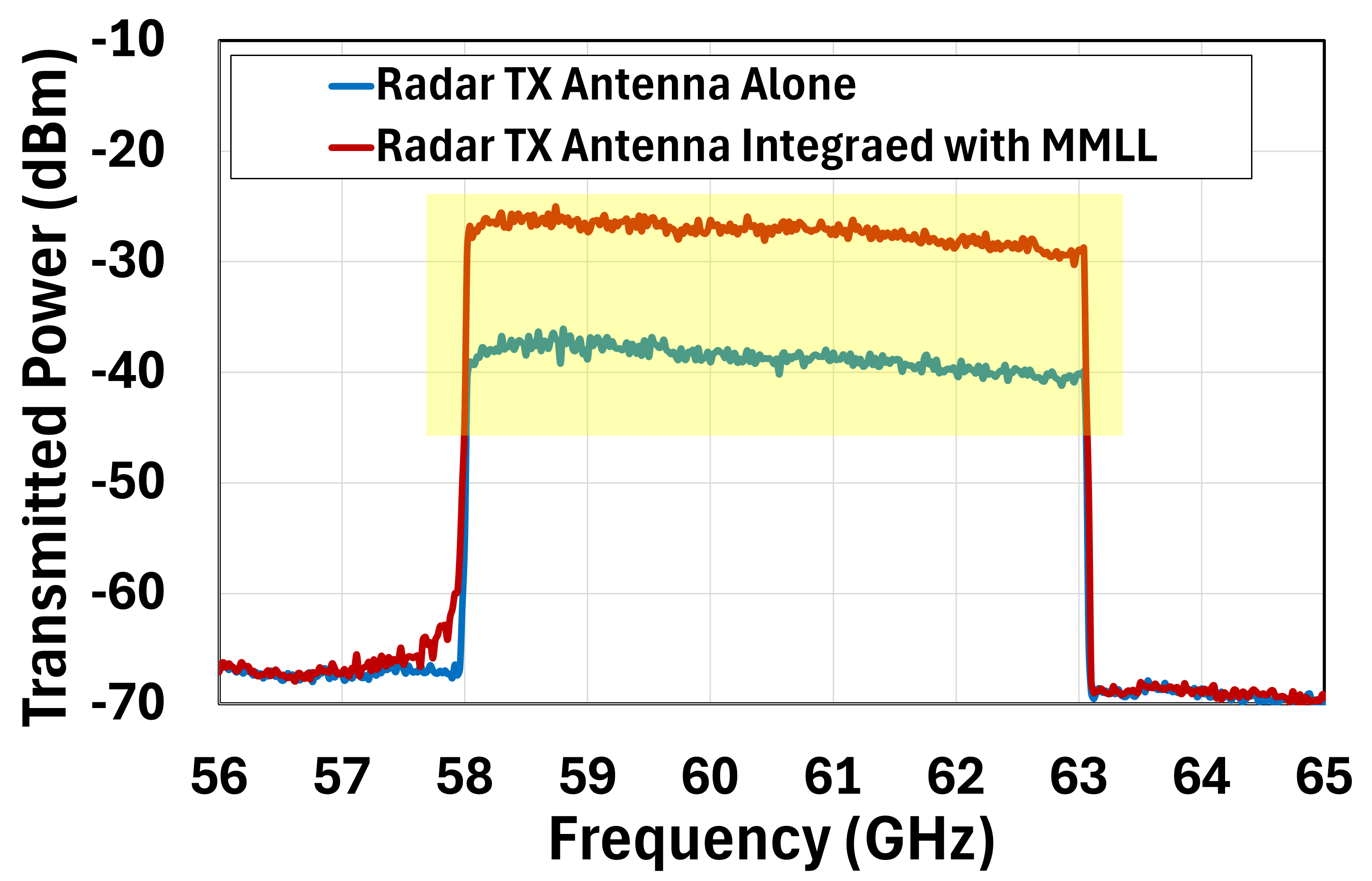}
  \caption{Measured transmitted power over 58–63 GHz, with and without the modified lens.}
  \label{fig7}
\end{figure}
%%%%%%%%%%%%%%%%%%%%%%%%%

In this work, the GRIN profile associated with the rotationally inserted dielectric rod is implemented using three discrete permittivity segments, as practical fabrication constraints prevent realizing a fully continuous gradient. Combined with the use of unit-cell dimensions of $0.4\lambda$, this discretization introduces phase errors that reduce the realized gain to approximately 20 dB. Although this gain reduction is not ideal and suggests opportunities for improved fabrication with higher spatial resolution in future implementations, the resulting performance remains satisfactory for the intended application. As illustrated in Fig.~\ref{fig6}(b) and Fig.~\ref{fig6}(c), full-wave simulations clearly highlight the fundamental difference between the proposed MMLL and a conventional GRIN Luneburg lens when driven by five spatially distributed bistatic radar modules. In Fig.~\ref{fig6}(b), the modified GRIN structure enables well-defined and consistently high-gain beams across all feeds, with realized-gain enhancements ranging from 13.5 dB for the central radar module (Radar \#3) down to 12.5 dB for the edge modules (Radar \#1 and Radar \#5). This slight and expected variation is a direct result of the engineered asymmetric rod-based GRIN distribution, which maintains proper phase correction and focusing for both central and off-axis feeds. In contrast, Fig.~\ref{fig6}(c) shows the simulated radiation patterns obtained when the five bistatic radar modules are integrated with a conventional GRIN Luneburg lens. Unlike the proposed MMLL, the classical GRIN profile fails to provide usable beam formation for off-axis radar feeds. The resulting patterns exhibit severe distortion, suppressed gain, and the absence of well-defined main lobes, with realized gains falling as low as –20 dB across elevation angles. No distinct or stable beam directions appear for any of the five radars, confirming that the classical GRIN-based focusing mechanism collapses under multi-feed excitation. This behavior is expected because the standard Luneburg profile is inherently optimized for a single on-axis feed, and does not include any mechanism to compensate for the lateral TX/RX displacement, feed asymmetry, or the mutual coupling introduced by closely spaced bistatic modules.

To assess the transmit-power enhancement of the MMLL in the experimental setup and to compare it directly with the simulation results, the TX antennas of the radar modules were used as the transmitting units, while a standard mm-wave horn antenna connected to a spectrum analyzer served as the receiver, as illustrated in Fig.~\ref{fig5}. The measurements were conducted across the full 58–63~GHz operating band of the Infineon BGT60TR13C radar modules to capture frequency-dependent variations in gain and radiation behavior. All measurements were performed inside a controlled anechoic chamber to eliminate environmental reflections, suppress multipath interference, and obtain high-fidelity power readings.

Corresponding measured results, depicted in Fig.~\ref{fig7}, confirm a closely matching gain improvement of approximately 12~dB in transmitted power across the desired frequency band. This strong correlation between simulation and experimental data validates the effectiveness of the proposed lens design in enhancing transmission performance. To comprehensively evaluate the spatial radiation performance, the receiving mm-wave horn antenna was mounted on a rotational stage and moved around the lens in a circular trajectory to capture the transmitted power at the designated angular positions of the five radar modules. Each radar unit, positioned at a fixed angular offset along the periphery of the modified Luneburg lens, generates a distinct high-gain beam corresponding to its assigned direction. The measured power profiles from the horn antenna confirm consistent gain enhancement across all five beams, demonstrating uniform energy distribution and stable angular coverage provided by the lens. The slight discrepancy between simulated and measured gain values is primarily attributed to fabrication tolerances, minor misalignment between radar and horn antenna positions, and material property variations at mm-wave frequencies.

To evaluate the spatial selectivity of the MMLL-assisted multi-radar system, the response of each radar module was examined across different angular sectors. When a target was positioned along the main beam direction of a given radar module, only that radar exhibited a strong return, while adjacent modules showed negligible variation. This behavior confirms that the proposed MMLL provides robust angular discrimination and feed-to-feed isolation, enabling independent multi-beam operation across all sensing directions.

%%%%%%%%%%%%%%%%%%%%%%%%%%%%%%%%%%%%%%%%%%%%%%%%%%%%%%%

\section{Sensitivity Analysis of the Multi-Beam GRIN-Based Radar System}

The performance and robustness of the proposed multi-beam radar architecture depend on a combination of electromagnetic, geometric, and structural factors that collectively determine tracking reliability and sensing fidelity. Although the system is designed around fixed-beam radars integrated with a GRIN-based lens, variations in antenna beamwidth, zone spacing, human-body angular extent, and the discretization level of the GRIN medium can all influence the overall sensing behavior. To quantify these effects, we present a detailed sensitivity analysis that examines the dominant contributors to system performance.

\subsection{MMLL Sensitivity to Unit-Cell Discretization}

The performance of a GRIN lens is fundamentally constrained by the ratio between the structural unit-cell size $L$ and the guided wavelength $\lambda_{g}$ inside the host dielectric. As the unit cell becomes electrically large, the effective-medium approximation breaks down, higher-order Floquet modes are excited, and phase errors accumulate, ultimately degrading the focusing performance. The analysis in \cite{wang2023high} demonstrates that the practical upper-frequency limit of a printed GRIN lens corresponds to the critical figure of merit presented in (9), beyond which the material can no longer be treated as a homogenized medium. 

%%%%%%%%%%%%%%%%%%%%%%%%%%%%%%
\begin{equation}
{
f_{\max}
=
\frac{1.4 \times c}{L\sqrt{\varepsilon_{r}}}
}
\label{eq:fmax}
\end{equation}
%%%%%%%%%%%%%%%%%%%%%%%%%

The guided wavelength within the $\varepsilon_{r}=2.8$ dielectric constrains the validity of the effective-index approximation. For a unit-cell size of $L = 0.4\lambda$, the corresponding electrical-size limit yields a maximum reliable operating frequency of approximately $62.8~\text{GHz}$. Since the radar operates from 58–63 GHz, the lens is driven near the boundary of this regime, where discretization of the continuous GRIN profile introduces quadratic phase errors \cite{ishimaru2017electromagnetic}. These deviations account for the reduced gain enhancement at higher frequencies and highlight the inherent trade-off between electromagnetic fidelity and the minimum manufacturable unit-cell dimension.

%%%%%%%%%%%%%%%%%%%%%%%%%%%%%%%%%%%%%%%%%%%%%%
\subsection{Angular Coverage Sensitivity Analysis}

\begin{figure}[t!]
  \centering
  \includegraphics[width=0.6\linewidth]{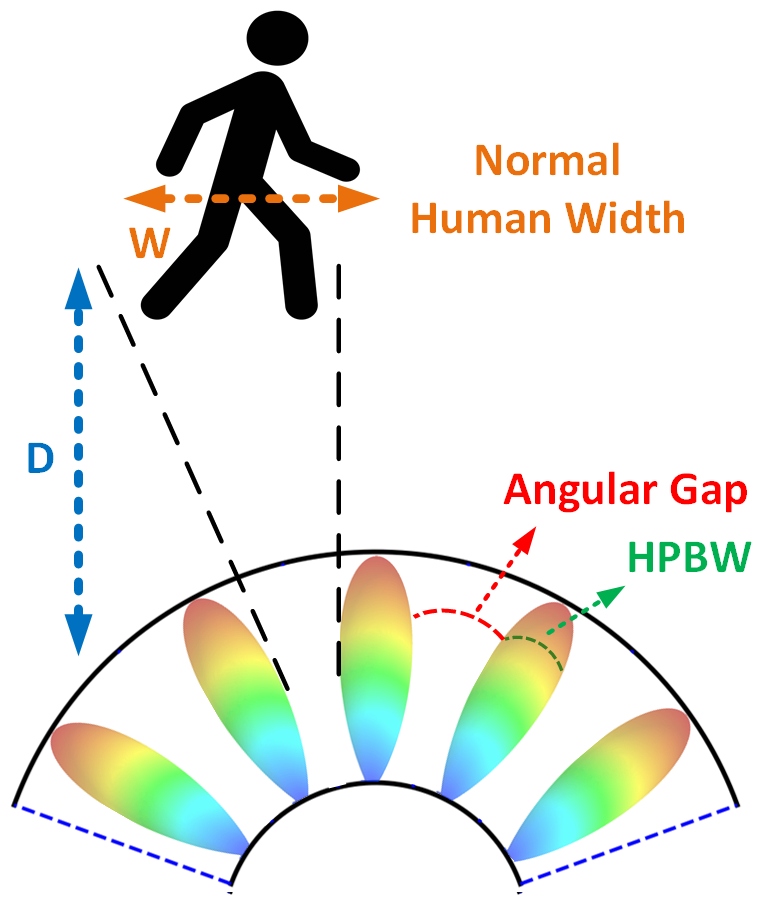}
  \caption{Geometric illustration of the angular-coverage model used in the proposed MMLL system.}
  \label{fig8}
\end{figure}

In the proposed radar-based tracking and fall-detection system, five high-gain antennas cover a $140^\circ$ azimuth sector with a nominal inter-beam spacing of $\Delta\theta = 28^\circ$. Each MMLL-integrated antenna produces a narrow half-power beamwidth (HPBW) of roughly $4^\circ$, as shown in Fig.~\ref{fig6}(a), which would traditionally imply a $24^\circ$ gap between adjacent $-3$~dB beam edges. However, unlike point-target angle-estimation systems, the present architecture (Fig.~\ref{fig8}) exploits the finite physical width of a human body and the short sensing range to achieve continuous, gap-free detection without requiring explicit beam overlap.

%%%%%%%%%%%%%%%%%%%%%
%%%%%%%%%%%%%%%%%%%%%%%%%

Considering $W$ as the typical left-to-right width of a human torso, anthropometric data place this value in the range of $0.45$--$0.55~\mathrm{m}$ for most adults. When a subject is located at a horizontal range $D$ in front of the radar cluster, the angular half-width subtended by their torso is,

\begin{equation}
\alpha = \arctan\left(\frac{W/2}{D}\right).
\end{equation}

The total angular width of the human body is therefore $2\alpha$. To ensure that a subject cannot be lost between adjacent beams during upright motion, the human angular width must be sufficient to cover the inter-beam gap. In the worst case, when the subject’s center lies exactly at the midpoint between two zones, the requirement for gap-free visibility is $\alpha \geq \frac{\theta_{\mathrm{gap}}}{2} = \frac{24^\circ}{2} = 12^\circ$. Using a representative human width of $W=0.50~\mathrm{m}$, the resulting distance is $D=1.18 m$.

This analysis shows that, for operating distances on the order of $1.0$--$1.2~\mathrm{m}$, the angular span of a standing human body (approximately $24^\circ$--$30^\circ$) is sufficient to overlap adjacent radar beams even though the antennas themselves do not overlap at the $-3$ dB level. Thus, the coverage continuity is guaranteed not by beam shape but by the target’s finite physical extent. 

On the other hand, the angular resolution of the proposed multi-beam radar system depends not only on the intrinsic antenna beamwidth but also on the apparent angular width of the human body, which varies significantly with distance. For an adult torso width of $W \approx 0.50~\mathrm{m}$, the apparent angular span of the body becomes significantly large at close distances ($R < 1.2~\mathrm{m}$). As a result, one individual tends to illuminate multiple adjacent beams simultaneously, and two people located near each other are likely to be perceived as a single merged return. As the target moves farther away, the apparent angular width decreases rapidly. In this far-field regime, the angular resolution is no longer limited by the human-body extent but is instead dominated by the beamwidth of the antennas. Thus, two individuals may be resolved as separate targets provided their angular separation exceeds approximately $4^\circ$, corresponding to a physical separation of about $0.7~\mathrm{m}$ at a distance of $10~\mathrm{m}$.

%%%%%%%%%%%%%%%%%%%%%%%%%%%%%%%%%%%%%%%%%%%%%%%%%

\section{Tracking System Analysis with MMLL Integrated Multi-Radar Architecture}

This work presents a radar-based tracking system for continuous, real-time patient monitoring, particularly suited for elderly individuals and those with mobility limitations. The system provides non-contact, privacy-preserving sensing without requiring wearable devices, offering a practical alternative to camera-based solutions. Real-time beam-tracking and signal-processing algorithms distinguish routine activities from critical events such as falls, with an automatic alert mechanism triggered when prolonged inactivity is detected. The compact setup operates on a standard laptop and features a modular architecture that supports multi-room or multi-patient deployment. Future extensions will incorporate AI-driven analytics for improved event classification and predictive healthcare monitoring.

%%%%%%%%%%%%%%%%%%%%%%%%%%%%%%%%%%%%%

\subsection{Multi-Radar Fusion System Overview}

In the proposed system, each radar module within the coordinated multi-radar framework simultaneously transmits and receives frequency-modulated continuous-wave (FMCW) signals to cover the designated scanning area. The receiver antennas capture the reflected echoes from the patient, measuring both the magnitude and phase of the backscattered signals. The strategic spatial arrangement of these radar units, guided by the focal characteristics of the modified Luneburg lens, enables angular and range estimation for effective beam tracking. Leveraging the inherent range-resolving capability of FMCW operation, the system continuously monitors subject motion and position across multiple locations, forming the foundation for real-time, non-contact tracking in indoor healthcare environments.

The fall-detection system is developed using the Infineon Radar SDK and five BGT60TR13C radar sensors, each configured to detect objects within a 2-m range from the antenna, as summarized in Table~\ref{tab1}. Although the radar datasheet specifies a maximum detection range of up to 15 m under ideal conditions, integrating the proposed Luneburg lens increases the effective range by approximately four times due to the enhanced antenna gain. Nevertheless, for fall-detection applications, a 2-m operational range is intentionally selected, as it aligns with typical indoor monitoring distances and ensures reliable high-resolution sensing in confined clinical or assisted-living environments.

%%%%%%%%%%%%%%%%%%%%%%%%%%%%%%%%%%%%%%%%%%%%%%%%
\begin{table}[!t]
\caption{BGT60TR13C Radar Parameters Used in the Proposed System (Detection Range up to 2 m)}
\label{tab1}
\centering
\small  % or \scriptsize for tighter spacing
\begin{tabular}{|c||c|}
\hline
\textbf{Parameter} & \textbf{Value} \\
\hline
Frame Repetition Time  & 0.05 s \\
\hline
Chirp Repetition Time  & 0.0007 s \\
\hline
Chirps per Frame       & 16 \\
\hline
Starting Frequency     & 58 GHz \\
\hline
Ending Frequency       & 63 GHz \\
\hline
Sample Rate            & 2.330 MHz \\
\hline
Samples per Chirp      & 128 \\
\hline
IF Gain                & 23 dB \\
\hline
\end{tabular}
\end{table}
%%%%%%%%%%%%%%%%%%%%%%%%%%%%%%%%%%%%%%%%%%%%%%%

Each radar unit includes one transmit and three receive antennas; however, in this implementation, only a single receiver, RX3, is activated. While using multiple receivers could improve azimuth estimation, it would significantly reduce presence detection resolution at short range. Therefore, the BGT60TR13C modules are operated in a simplified 1TX/1RX mode to streamline processing and maintain reliable amplitude-based detection. This configuration is sufficient for real-time motion and fall detection while reducing data complexity. Future work will extend the system to leverage the full SIMO capability of each radar module within the modified Luneburg lens sectors to enhance angular resolution and target localization accuracy.

The system architecture is divided into two main components: a back-end for radar data acquisition and processing, and a front-end for visualization and user interaction. To support simultaneous operation of all five FMCW radars, the back-end uses a hybrid multi-threading and multi-processing design in Python, allowing each radar to run in its own process for true parallel data capture. The front-end provides real-time visualization and instant detection updates, resulting in a responsive software framework tailored to multi-beam sensing. 

%%%%%%%%%%%%%%%%%%%%%%%%%%%%%%%%%%%%%%%%%%
\subsection{Backend Architecture: Multi-Radar Signal Processing and Fusion Framework}

The backend architecture serves as the computational core of the proposed fall-detection and tracking system, handling real-time signal processing, radar synchronization, and multi-sensor data fusion. Each radar operates as an independent node within its own child process, continuously streaming amplitude and phase data to the central processor. A Python-based multiprocessing framework assigns a dedicated process to each radar module, enabling parallel data acquisition, pre-processing, and frame management to ensure low latency and maintain temporal alignment across all sensors. The processed outputs, representing detection states and reflection strengths, are fused in a centralized decision layer that correlates spatial information from all modules to generate a unified occupancy assessment. Binary detection values and associated metadata are transferred to the front-end through a shared multiprocessing queue, while a dedicated communication thread maintains reliable synchronization between the processing backend and the graphical interface, ensuring seamless real-time monitoring.

Since each radar module exhibits a distinct baseline amplitude response due to hardware tolerances and environmental variations, the backend performs an initial calibration routine in which incoming amplitude values are averaged over a defined number of iterations (typically 100) per radar. This averaged baseline serves as a reference for subsequent detection decisions. A target is identified as present when the measured amplitude exceeds the baseline by a predefined offset. This offset, determined empirically through extensive measurement trials, was set to 0.75~dB to achieve an optimal balance between detection sensitivity and false-alarm suppression. This calibration approach ensures reliable human presence detection while maintaining robustness against noise fluctuations and minor environmental changes.

To acquire amplitude data from each radar module, the main process maintains a shared queue that communicates with all active child processes. Each child process, upon completing the processing of a radar frame, pushes the corresponding amplitude data along with its associated radar \textit{uuid} into the shared queue. The main process continuously polls this queue to retrieve the incoming data and determines, for each radar, whether the measured amplitude exceeds the predefined threshold. Based on this comparison, the system classifies the state as either the presence or absence of a person in front of the radar.

%%%%%%%%%%%%%%%%%%%%%%%%%%%%%%%%%%%%%%
\subsection{Frontend Architecture: Coordinated Multi-Radar Sensing and Data Acquisition}

\begin{figure}[!t]
\centering

\subfloat[]{\includegraphics[width=0.7\columnwidth]{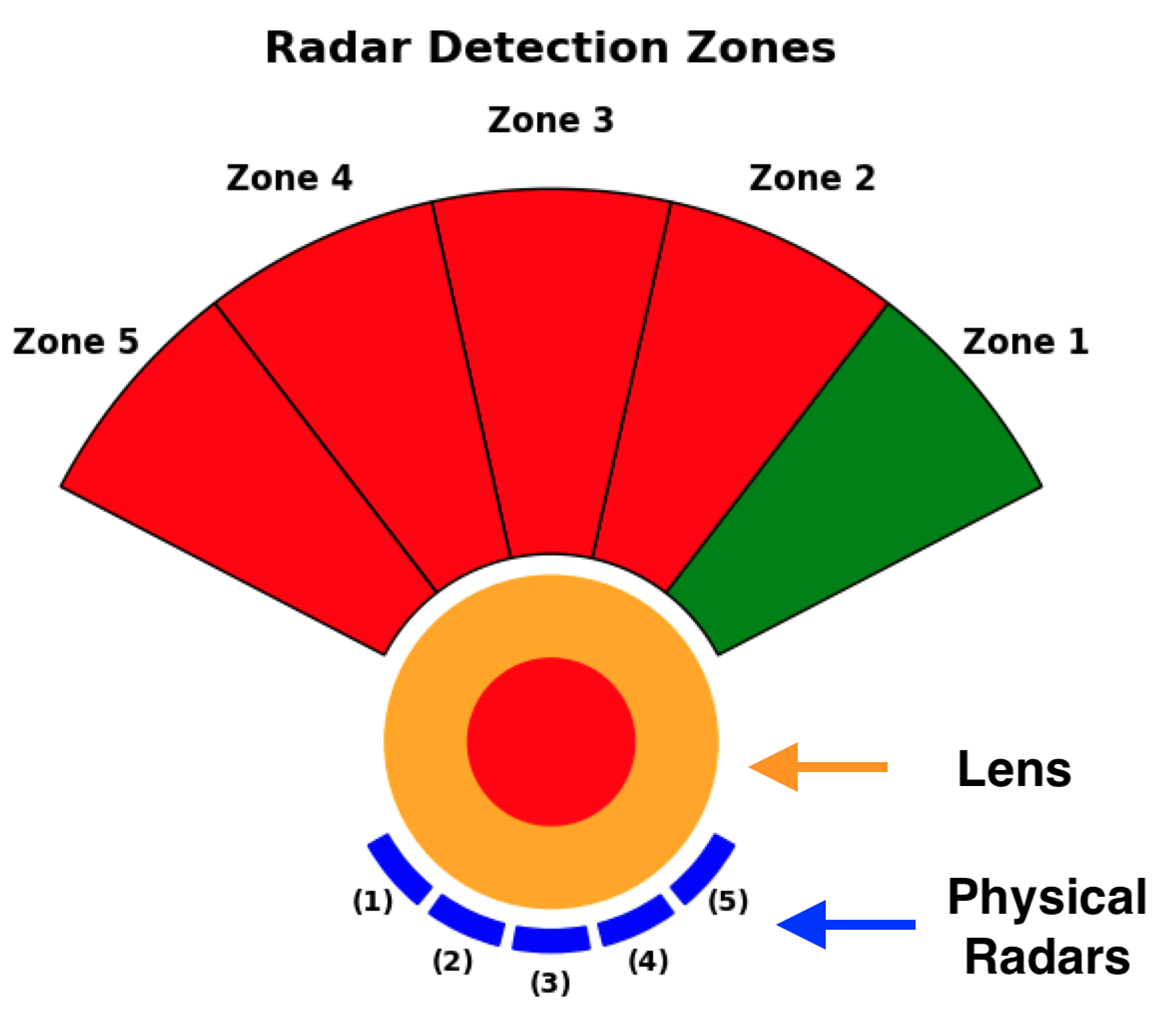}}\label{fig9a}\hfill
\subfloat[]{\includegraphics[width=\columnwidth]{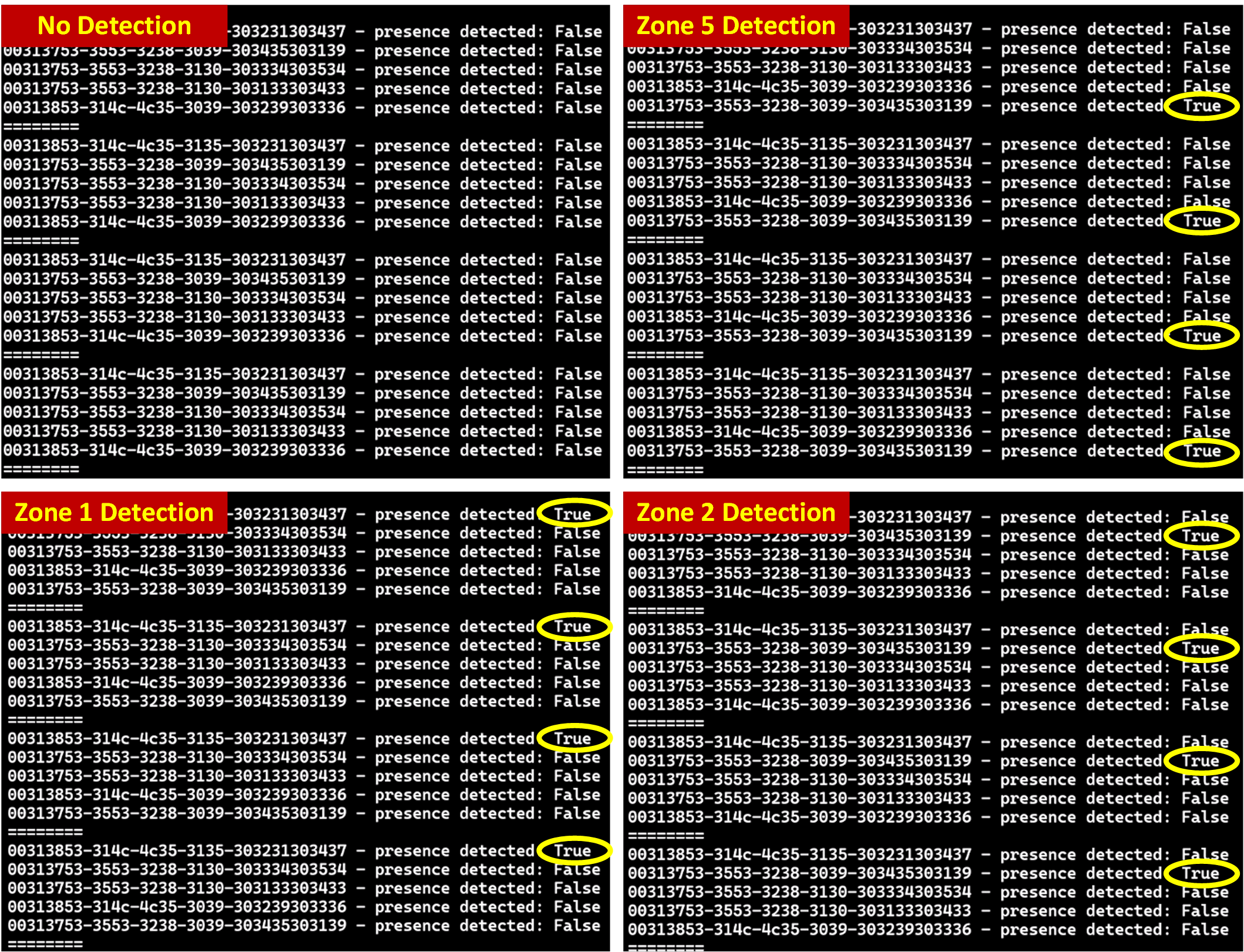}}\label{fig9b}

\caption{(a) Active detection GUI showing real-time radar zone updates and visualization of detected person. (b) Python-generated presence-detection outputs from the MMLL system, showing that the radar corresponding to the target’s angular zone reports “True” while all others remain “False.}
\label{fig9}
\vspace{-1mm}
\end{figure}

The frontend interface of the proposed system is designed to provide an intuitive, real-time visualization of the multi-radar sensing environment. Developed in Python for dynamic rendering, the interface employs a polar-coordinate framework that mirrors the physical arrangement of the radar modules around the modified Luneburg lens. The front-end retrieves processed data from the shared queue and provides real-time visual feedback through an integrated graphical user interface (GUI), as illustrated in Fig. \ref{fig9}(a). The GUI operates in continuous synchronization with the backend, refreshing dynamically as new JSON data streams are received from each radar module. Each zone’s state is updated in real time based on its corresponding radar’s unique identifier (\texttt{UUID}) and associated \texttt{detect} variable, enabling instantaneous visualization of presence or absence within the monitored regions. 

%%%%%%%%%%%%%%%%%%%%%%%%%%%%%%%%%%%%%%%%%%%%%%%%%
%%%%%%%%%%%%%%%%%%%%%%%%%%%%%%%%%%%%%%%%%%%%%%%

Each radar module is mapped to a predefined spatial sector, allowing zone-specific representation of activity within the monitored area, while a central circular heatmap illustrates the focal area of the lens where beam convergence occurs. The radar modules are depicted as blue arcs along the lower periphery of the display. Zone colors are continuously updated to reflect detection status, \textcolor{green}{green} indicates human presence within the corresponding zone, whereas \textcolor{red}{red} signifies absence. Beyond color-coded indicators, the interface also provides textual feedback at key detection events, enhancing situational awareness and facilitating rapid operator response, as shown in Fig. \ref{fig9}(b). This real-time visual feedback enhances situational awareness, allowing rapid interpretation of patient location and movement.

As detection results arrive, the system cross-references each radar’s identifier with its assigned zone and dynamically updates the interface, flagging a sector as occupied when the corresponding radar reports an amplitude above the detection threshold. This real-time pipeline continuously acquires radar data, performs presence detection using amplitude and phase analysis, and executes a multistage fall-detection algorithm, with the GUI delivering both visual and auditory alerts.

The intentional adjacency of radar coverage zones enables seamless handoff between sensors, maintaining uninterrupted tracking as the patient moves through different areas. The fall-detection logic follows a four-stage cyclic process embedded within the main GUI control loop. In the \textit{Initial Detection} stage, the system continuously scans all five zones until a subject is detected, recording the detection zone as Zone~X to initiate tracking. During the \textit{Monitoring for Disappearance} stage, the system verifies signal persistence within Zone~X; if detection ceases, it transitions to the fallback stage. The \textit{Fallback Check and Timer} stage activates an alert timer upon signal loss—20~s for edge zones (1 and 5) and 10~s for middle zones (2–4), while monitoring adjacent zones (X-1,X,X+1) for signal reappearance. Finally, in the \textit{Reappearance or Alert} stage, detection within the timeout window updates the tracking zone and normal operation resumes; otherwise, when the timer expires, the system triggers an alert, issues a GUI warning, and dispatches an emergency notification to the designated caregiver.

The modular software architecture ensures scalability, allowing straightforward integration of additional radar modalities such as micro-Doppler signal analysis or deep-learning–based posture and activity classification in future system iterations.

%%%%%%%%%%%%%%%%%%%%%%%%%%%%%%%%%%%%%%%%%%%%%%%%%%

\section{System Implementation and Validation Framework}

\begin{figure}[!t]
\centering

\subfloat[]{\includegraphics[width=\columnwidth]{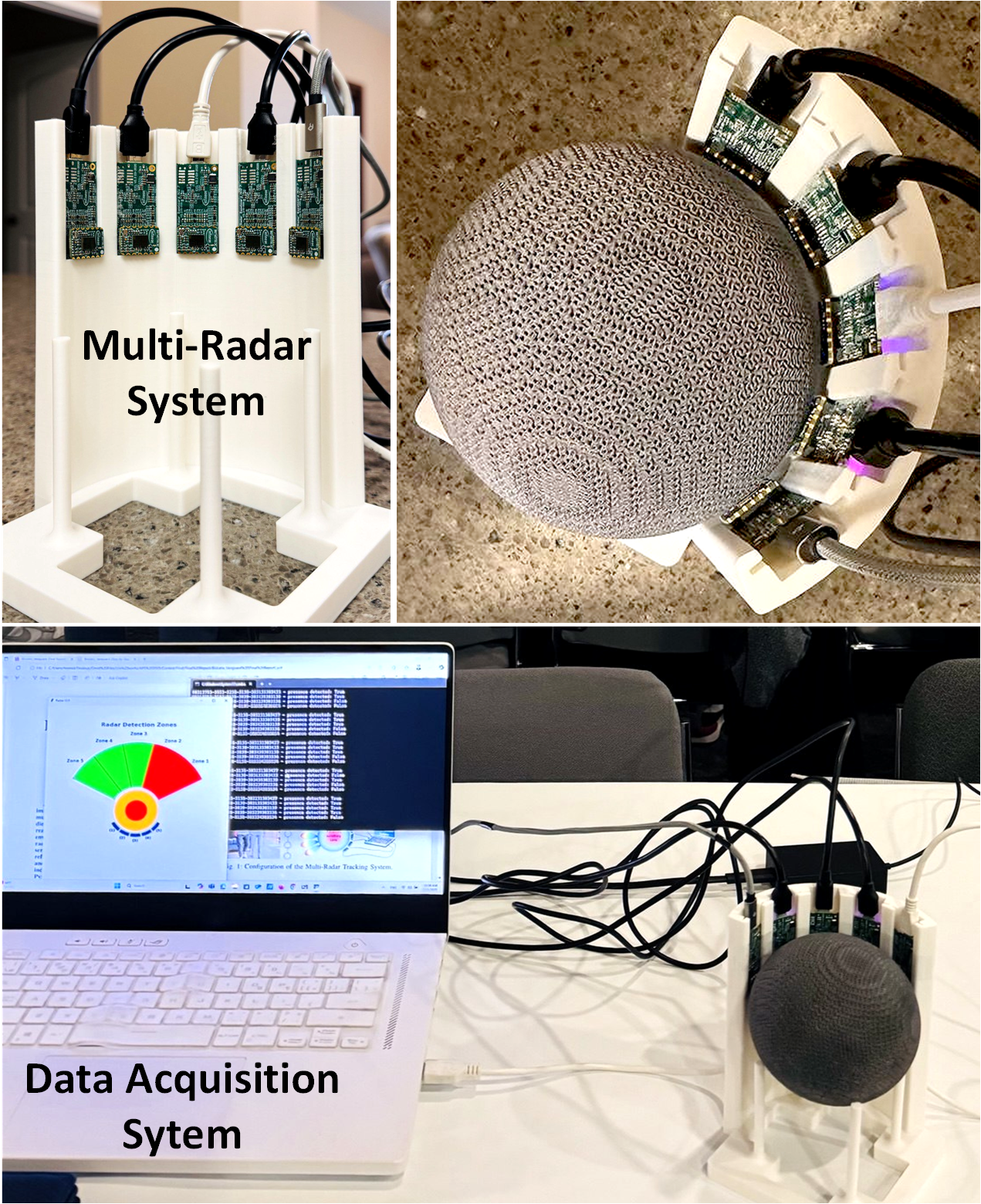}}\label{fig10a}\hfill
\subfloat[]{\includegraphics[width=\columnwidth]{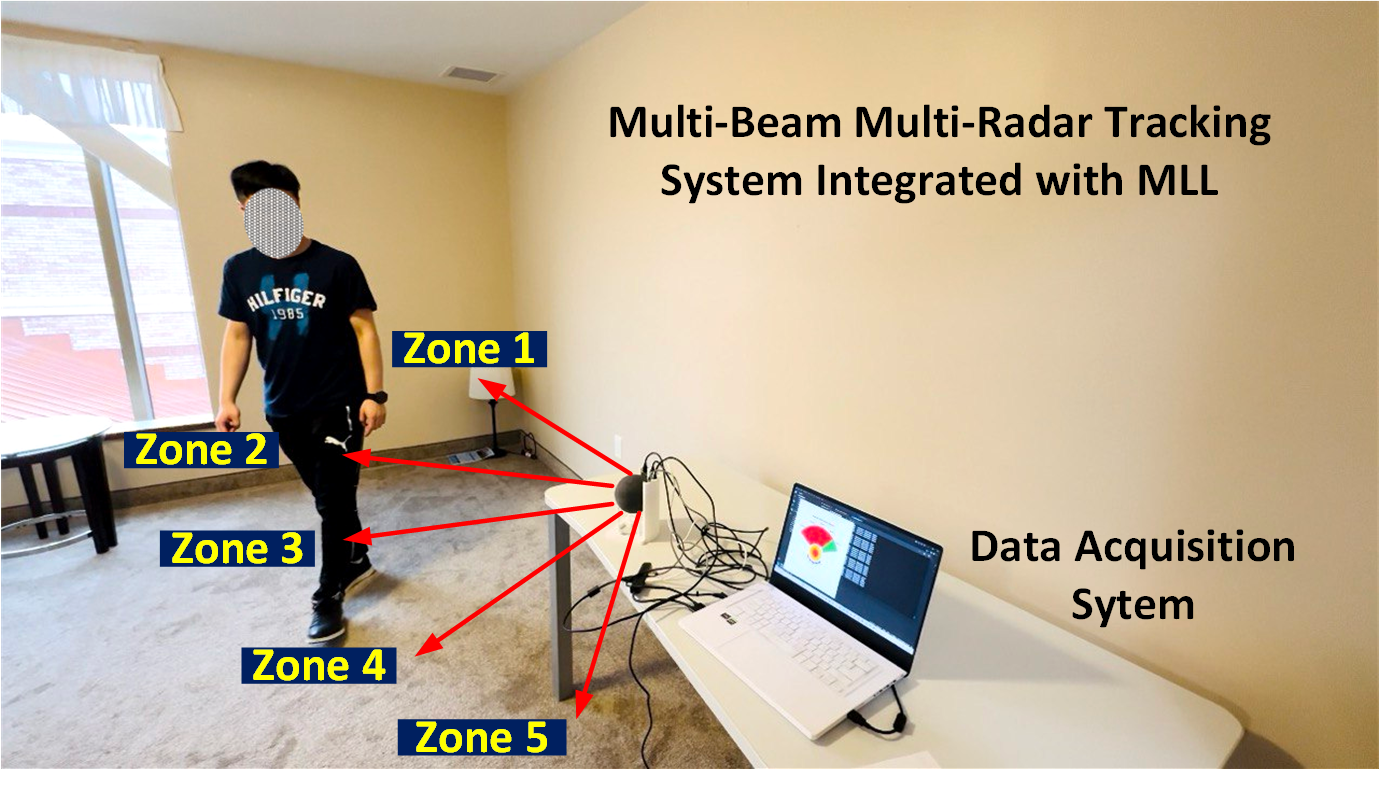}}\label{fig10b}

\caption{Prototype and experimental setup of the proposed multi-radar system integrated with the MLL. (a) Five Infineon BGT60TR13C radar modules are mounted on a 3D-printed semicircular fixture around the MLL, forming the core of the multi-radar array integrated with a data-acquisition unit for real-time signal processing and visualization. (b) The five radar modules define overlapping detection zones (Zones 1–5) within the monitored area, enabling real-time tracking.}
\label{fig10}
\end{figure}

This section presents the complete implementation and experimental validation of the proposed multi-radar sensing system integrated with the modified GRIN Luneburg lens (MMLL). Following the design and characterization of the MMLL and the development of the coordinated multi-radar signal-processing framework, the system was assembled into a fully functional prototype for real-time evaluation. The experimental setup was designed to assess the system’s ability to detect and track human movement across defined spatial zones and to reliably identify fall events. Measurements were conducted using real human motion scenarios to evaluate detection accuracy, continuity of zone transitions, and the responsiveness of the alert mechanism under realistic operating conditions. The complete hardware realization and experimental setup are shown in Fig.~\ref{fig10}.

As shown in Fig. \ref{fig10}(a), the prototype system comprises five Infineon BGT60TR13C radar modules integrated with the MMLL and mounted on a custom 3D-printed fixture, forming five partially overlapping detection zones that correspond to distinct spatial sectors within the environment. The fixture ensures precise angular alignment, structural rigidity, and stable mechanical support for each radar while maintaining the intended semicircular geometry for optimal spatial coverage. The radars are interfaced with a central data acquisition unit via USB connections, enabling synchronized control and real-time data streaming through the Python-based backend.

%%%%%%%%%%%%%%%%%%%%%%%%%%%%%%%%%%%%%%%%%%%%%%%%%

%%%%%%%%%%%%%%%%%%%%%%%%%%%%%%%%%%%%%%%%%%%%%%%

%%%%%%%%%%%%%%%%%%%%%
\begin{figure*}[h]
  \centering
  \includegraphics[width=0.7\linewidth]{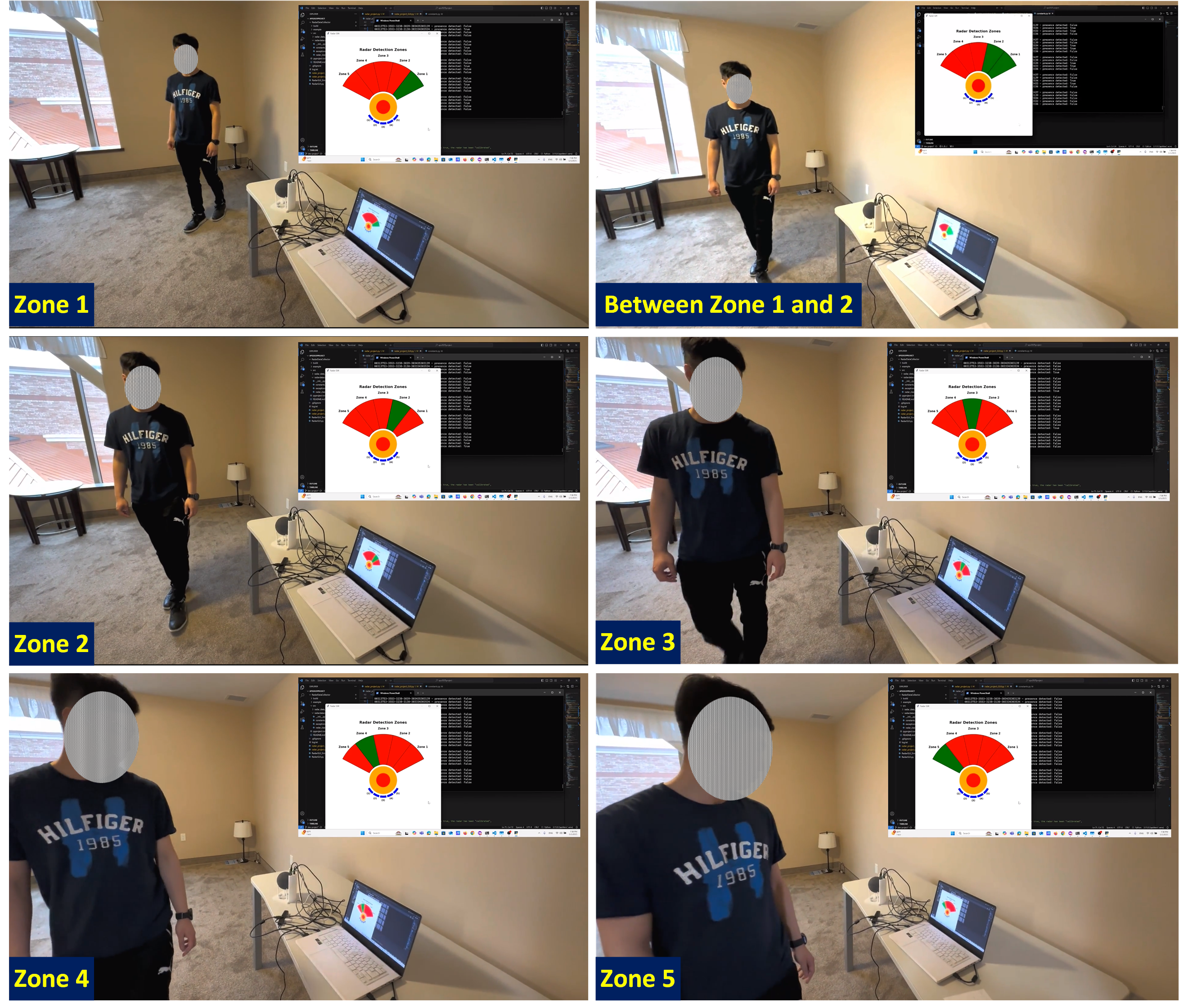}
  \caption{Real-time tracking performance of the proposed multi-radar system integrated with the MLL. Sequential frames illustrate a human subject moving from Zone 1 to Zone 5, demonstrating continuous detection across adjacent sensing zones.}
  \label{fig11}
  \vspace{3mm}
\end{figure*}
%%%%%%%%%%%%%%%%%%%%%%%%%
%%%%%%%%%%%%%%%%%%%%%
\begin{figure*}[h]
  \centering
  \includegraphics[width=0.7\linewidth]{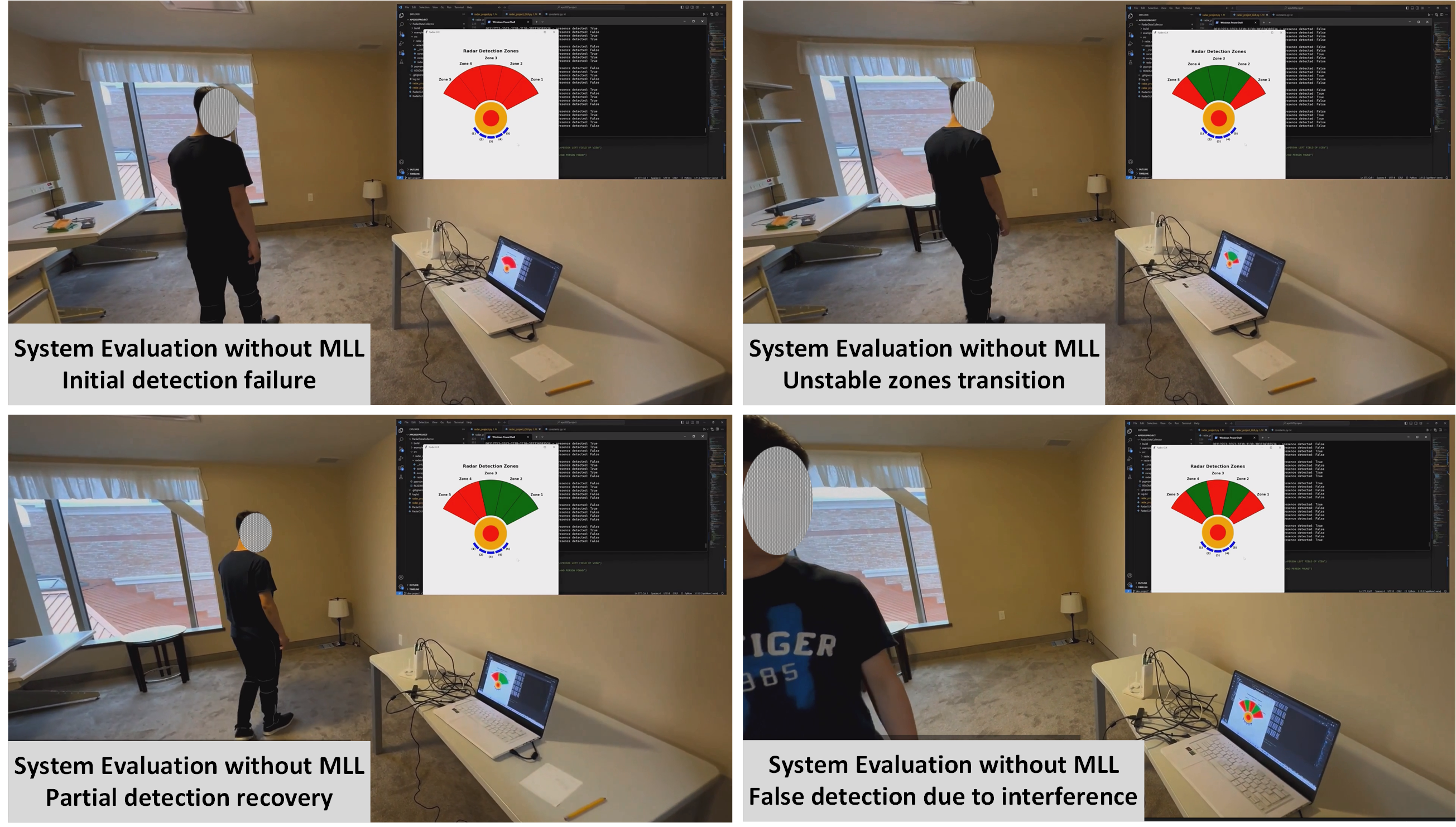}
  \caption{System performance evaluation without the modified GRIN Luneburg lens.}
  \label{fig12}
\end{figure*}
%%%%%%%%%%%%%%%%%%%%%%%%%

The experimental setup used for performance validation is depicted in Fig. \ref{fig10}(b). The measurements were conducted in a real healthcare environment at the \textit{Schlegel–UW Research Institute for Aging (RIA), Waterloo, ON, Canada}, to demonstrate the practical applicability of the system. The assembled MMLL-based multi-radar system was positioned at one end of the room and connected to a laptop running the real-time data acquisition and visualization software. Five spatially defined and partially overlapping zones (Zones 1–5) were assigned to the respective radar fields of view. During measurements, a human subject moved naturally through these zones and performed controlled activities, including walking, sitting, and simulated falls, to evaluate the system’s responsiveness. The GUI displayed the corresponding zone activations, while the backend monitored signal continuity and automatically triggered alerts when the subject’s signal disappeared for longer than the predefined threshold. This configuration validates the proposed system’s capability to achieve seamless multi-zone tracking, autonomous fall detection, and robust real-time operation under realistic indoor conditions.

Fig. \ref{fig11} demonstrates the real-time operation of the multi-radar tracking system integrated with the MMLL during human motion across the defined spatial zones. The figure presents sequential snapshots captured during a continuous walking experiment, showing both the subject’s position relative to the radar array and the corresponding zone detection output displayed on the graphical user interface (GUI). When the subject is located directly in front of a radar, the corresponding zone on the GUI becomes highlighted, indicating active detection. For instance, the subject is within Zone 1, followed by a transition to the intermediate position between Zones 1 and 2, and subsequently to Zones 2, 3, 4, and 5. This sequence validates the system’s ability to continuously track the subject’s movement across the overlapping radar coverage regions with smooth and reliable zone-to-zone transitions.

Throughout the experiment, the backend software continuously streamed amplitude data from all five radars to the GUI, which visualized detection results in real time. Notably, even in transition states, the system successfully maintained continuity of detection without abrupt signal loss or false triggers. Fig. \ref{fig12} presents the evaluation of the proposed multi-radar tracking system without integration of the modified Luneburg lens, highlighting the system’s degraded spatial coverage and increased vulnerability to environmental interference. The four frames capture representative moments during human motion within the same test environment used in Fig. \ref{fig11}, enabling a direct qualitative comparison with the MLL-assisted configuration. In the top-left frame, the subject is positioned near Zone 1, yet the GUI display indicates no active detection, confirming that the narrow beamwidth of the standalone radars prevents proper off-axis target sensing. In the top-right frame, as the subject transitions between Zones 1 and 2, intermittent responses appear on the GUI; however, these are inconsistent and fail to provide a stable tracking indication due to insufficient beam overlap between adjacent radars. When the subject moves closer to the center region (bottom-left frame, Zone 3), one radar begins to detect motion, activating a single sector on the GUI. This confirms that the system retains limited detection capability only along the direct line of sight of individual radar units, without the benefit of beam steering or gain enhancement provided by the MLL. Finally, the bottom-right frame shows a clear false detection event: although no subject is present within the radar’s field of view, the GUI erroneously reports activity. This false alarm results from multipath reflections and cross-interference between closely spaced radars, which become significant in the absence of the lens’s focusing and spatial filtering effects.

These results collectively demonstrate that the removal of the MLL leads to a marked reduction in detection uniformity and accuracy. Without the lens, the five radar units operate as isolated sensors, leaving blind regions between coverage sectors and amplifying multipath interference. Quantitatively, the system exhibited a 50\% reduction in consistent zone detection and significantly increase in false-alarm probability during repeated trials. In contrast, the MLL-assisted configuration maintained continuous detection across all five zones with negligible overlap ambiguity and near-zero false positives. The lens not only broadens the effective field of view but also mitigates unwanted reflections, stabilizing system performance and ensuring precise zone-to-zone tracking continuity.

While the proposed MMLL-assisted multi-radar system demonstrates strong performance in gain enhancement, coverage, and detection reliability, several practical challenges remain for future optimization. A primary limitation arises from the physical dimensions and housing of the Infineon BGT60TR13C radar modules, which restrict how closely the sensors can be positioned around the lens. This constraint limits the achievable angular resolution and prevents the addition of more radar units that could otherwise improve sensitivity and enable finer spatial discrimination. Importantly, this limitation is imposed by the radar hardware rather than the lens itself, as the modified GRIN Luneburg architecture inherently supports an arbitrary number of sensing elements positioned along its surface. The size of the modules also impacts the ability to maintain precise beam adjacency between neighboring radar zones; small mechanical misalignments in fixture geometry or mounting angles can shift the intersection of the half-power beamwidth regions and influence tracking continuity. Future work could extend the presented permittivity distribution over the entire spherical surface, rather than only radially, to enable sensor placement across a full hemisphere for 3D scanning. Another challenge arises from the discretization of the GRIN structure. To preserve permittivity continuity and minimize phase errors, the modified Luneburg lens ideally requires the smallest possible unit-cell dimension. In practice, the minimum feature size of the 3D-printing process limits the unit cell in this prototype to approximately $0.4\lambda$ at 60 GHz. At this scale, the effective-medium approximation begins to degrade, introducing discretization-induced phase deviations that reduce the achievable gain and focusing accuracy.

%%%%%%%%%%%%%%%%%%%%%%%%%%%%%%%%%%%%
\section{Conclusion}

This work introduced a multi-radar modified gradient-index Luneburg lens (MMLL) architecture designed for integration with mm-wave bistatic radar modules, enabling multi-beam, wide-angle sensing without mechanical or electronic beam steering. By embedding a rotationally symmetric, rod-based anisotropic dielectric region within a modified Luneburg permittivity distribution, the proposed lens supports the simultaneous formation of multiple high-gain beams, each naturally aligned with its corresponding radar module positioned along the lens periphery. Experimental validation demonstrated that each radar integrated with the MMLL achieved approximately 12 dB of realized antenna-gain enhancement per path, corresponding to a 24 dB two-way link-budget improvement, that enabled up to a fourfold extension in detection range. The multi-radar system provided uniform high-gain performance across five angular sectors spanning a total of 140$^\circ$, enabling robust multi-directional sensing within a compact form factor. A real-time prototype demonstrated effective fall detection and motion tracking across the five spatial zones, supported by an intuitive graphical user interface for visualization and autonomous alert generation. The 3D-printed implementation of the MMLL, combined with the modular radar architecture, offers a low-cost, scalable solution suitable for healthcare, assisted-living, and smart-environment monitoring. Future extensions may incorporate additional radar modules and vertical stacking to achieve full 3D coverage, further improving spatial awareness and reducing misclassification between falls and routine movements.

%%%%%%%%%%%%%%%%%%%%%%%%%%%%%%%%%%%%%%%%%
\section*{Acknowledgment}

The authors would like to acknowledge the support of the Wireless Sensors and Devices Lab (WSDL) at the University of Waterloo. Special thanks are extended to Fortify Inc. for their contribution to the fabrication process and technical assistance throughout the development of this work.

%%%%%%%%%%%%%%%%%%%%%%%%%%%%%%%%%%%%%%%%%%%%%%%%%

\bibliographystyle{IEEEtran}
\bibliography{Ref}

\vspace{-3em}

\begin{IEEEbiography}[{\includegraphics[width=1.1in,height=1.3in,clip,keepaspectratio]{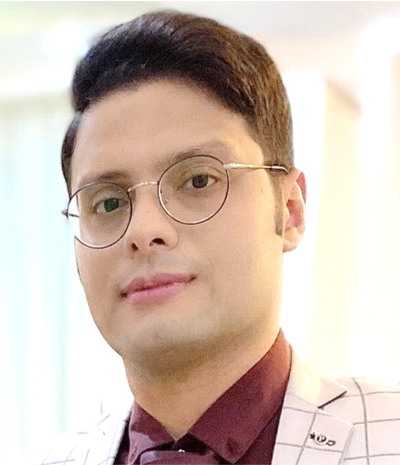}}]{Mohammad Omid Bagheri } (Member, IEEE) received the B.Sc. degree in Electrical Engineering from the Amirkabir University of Technology, Tafresh Branch, Iran, in 2012, the M.Sc. degree in Telecommunications Engineering from Shahed University, Tehran, Iran, in 2016, and the Ph.D. degrees in Telecommunications Engineering from Shahed University, Tehran, Iran, in 2021, and in Antennas, Microwaves, and Wave Optics from the University of Waterloo, ON, Canada, in 2025. He is currently a Postdoctoral Fellow with the Department of Electrical and Computer Engineering, University of Waterloo. Dr. Bagheri’s research has resulted in high-impact publications in Nature Communications, IEEE journals and conferences, and received several prestigious awards in IEEE APS and MTT-S communities. His work on non-invasive blood glucose monitoring for next-generation smartwatches was featured by CTV News Canada. His research interests include microwave imaging, biomedical sensing, array antennas, metasurfaces, and 3D-printed dielectric lenses.
\end{IEEEbiography}

\vspace{-3em}

\begin{IEEEbiography}[{\includegraphics[width=1in,height=1.25in,clip,keepaspectratio]{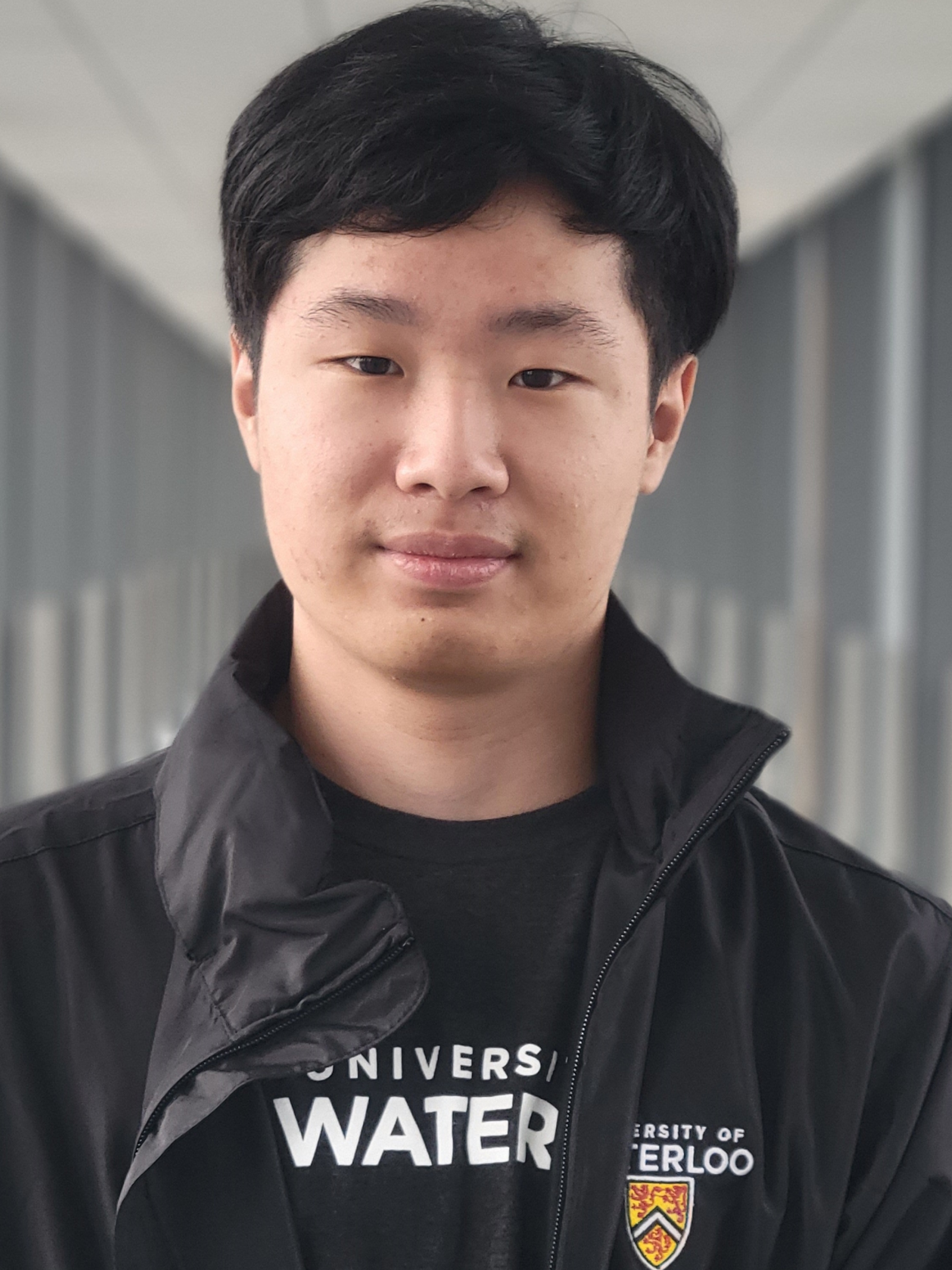}}]{Justin Chow} is currently pursuing his B.ASc in Computer Engineering, with a specialization in Signal Processing and Communications, at the University of Waterloo, Waterloo, Canada. He has been affiliated with the Wireless Sensors and Devices Lab at the University of Waterloo as an undergraduate researcher, contributing to research in radar systems and applications of 5G communications. With experiences in both software and embedded programming, he is eager to explore new research opportunities and drive innovation beyond his undergraduate studies.
\end{IEEEbiography}

\vspace{-3em}

\begin{IEEEbiography}[{\includegraphics[width=1in,height=1.4in,clip,keepaspectratio]{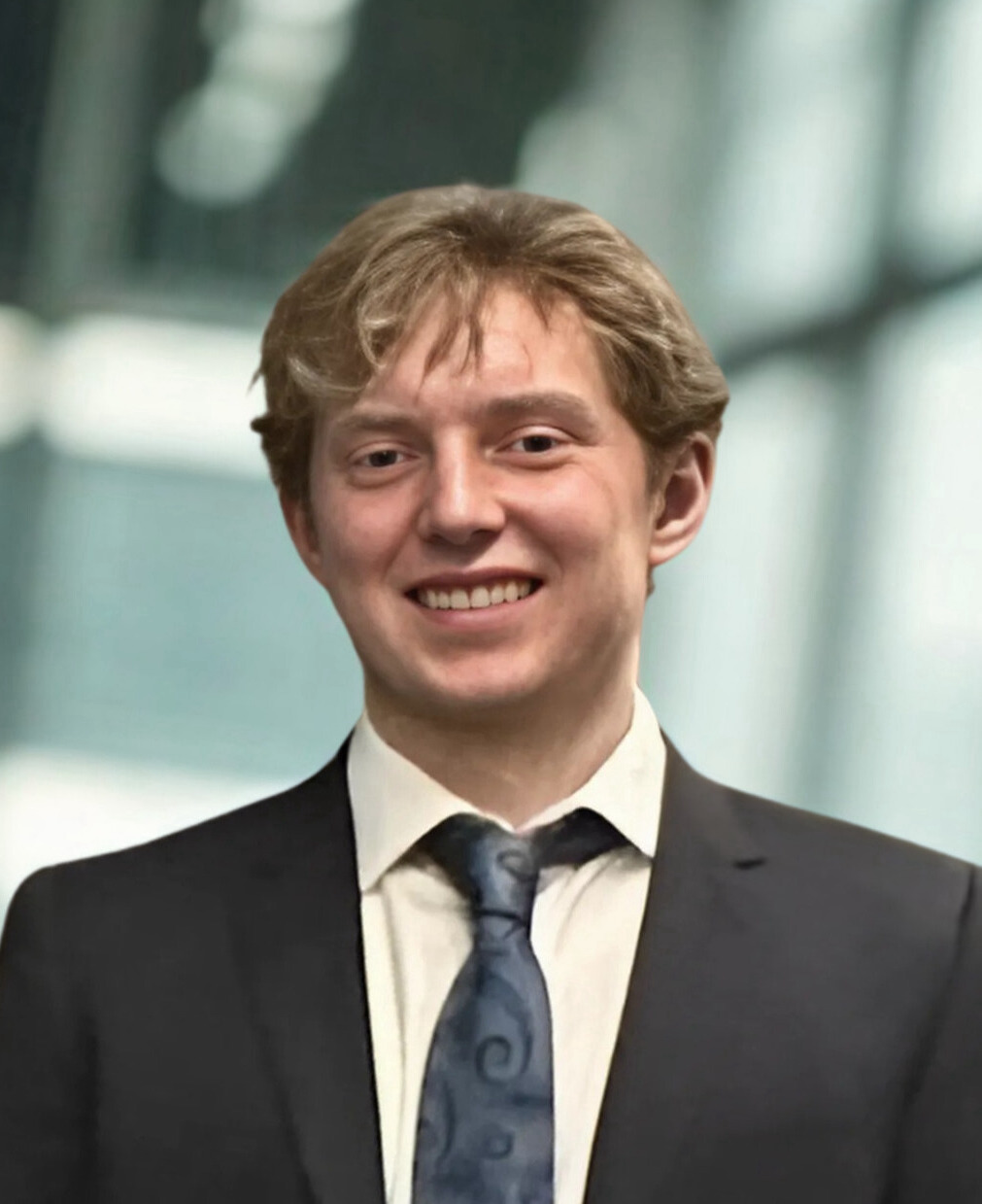}}]{Joshua Visser}
Received the B.Eng. degree in Biomedical Engineering from the University of Guelph in Spring 2025. He has been affiliated with the Wireless Sensors and Devices Laboratory at the University of Waterloo as an Undergraduate Researcher. His research focuses on radar-based monitoring systems for long-term care environments, integrating deep learning, signal processing, and embedded hardware design. He has experience in R\&D and hardware engineering, with roles at Evertz Microsystems and Accelovant Technologies.
\end{IEEEbiography}

\vspace{-3em}

\begin{IEEEbiography}[{\includegraphics[width=1in,height=1.25in,clip,keepaspectratio]{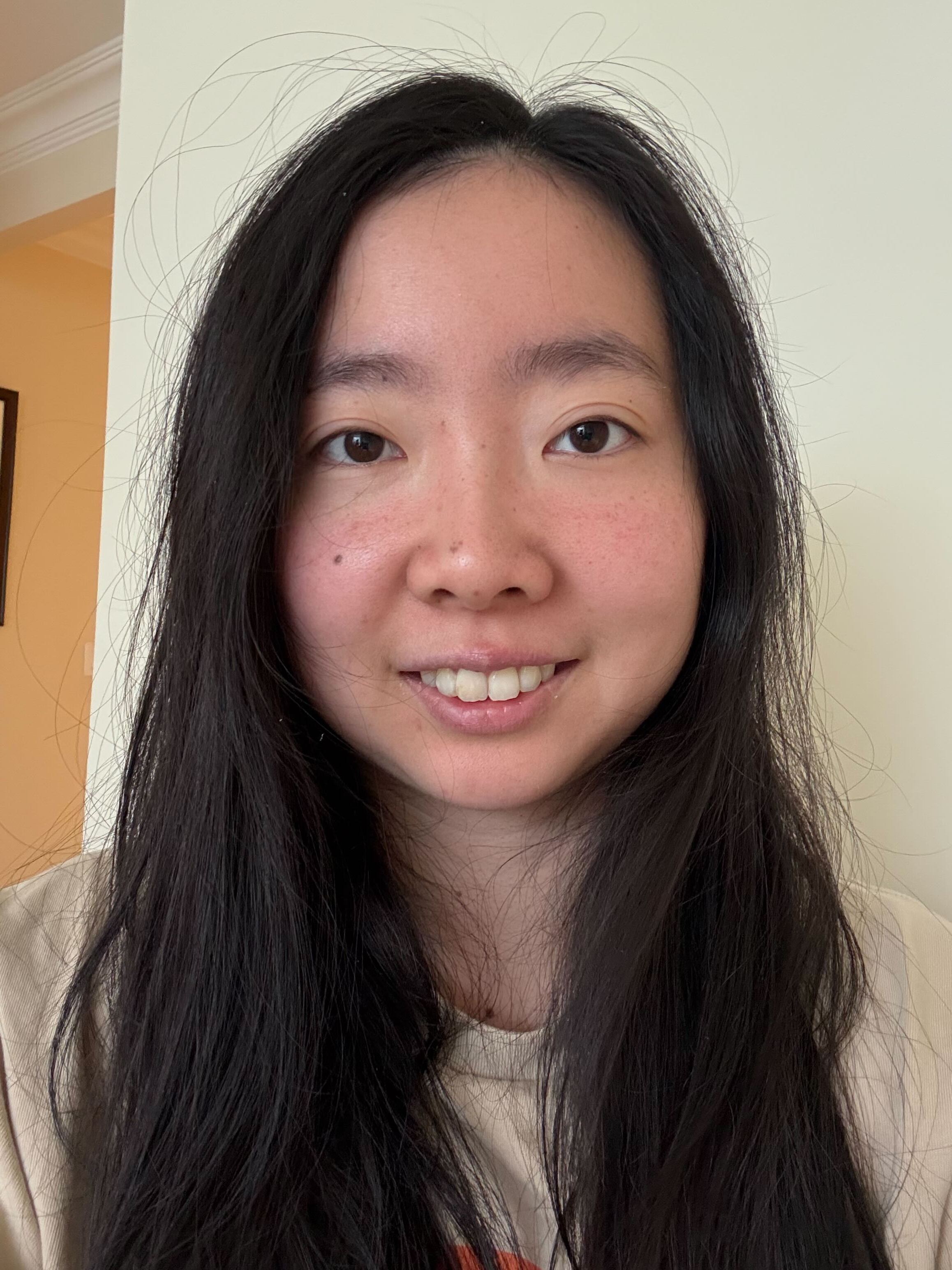}}]{Veronica Leong} (Graduate Student Member, IEEE) is currently pursuing the M.A.Sc. degree in Electrical and Computer Engineering at the University of Waterloo, Waterloo, ON, Canada, where she also received the B.A.Sc. degree in Mechanical Engineering in 2024. Her research focuses on radar-based biomedical sensing, including multimodal sensor fusion for non-invasive glucose monitoring and the development of millimeter-wave radar systems for healthcare applications. She combines expertise in mechanical design and RF systems to advance user-friendly and accurate biomedical monitoring technologies. 
\end{IEEEbiography}

\vspace{-3em}

\begin{IEEEbiography}[{\includegraphics[width=1in,height=1.25in,clip,keepaspectratio]{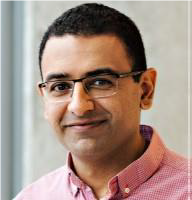}}]{George Shaker} (S IEEE 1997, SM IEEE 2015) is the Director of the Wireless Sensors and Devices Laboratory at the University of Waterloo–Schlegel Research Institute for Aging and serves as an Adjunct Professor in both the Department of Electrical and Computer Engineering and the Department of Mechanical and Mechatronics Engineering at the University of Waterloo. He was previously an NSERC scholar at the Georgia Institute of Technology and held senior engineering roles at Research In Motion (BlackBerry). With nearly twenty years of industry experience and over 8 years in academia, his work focuses on wireless sensor systems for healthcare, automotive, and UAVs. He has over 120 publications, 35 patents/patent applications, and multiple awards recognizing his academic and industrial contributions.
\end{IEEEbiography}

\end{document}